\documentclass{emulateapj}

\newcommand\Chandra{\textit{Chandra}}
\newcommand\asca{{\small {\it {ASCA}}}}
\newcommand\hmxb{{\small HMXB}}
\newcommand\ccd{{\small CCD}}
\newcommand\rrc{{\small RRC}}
\newcommand\hetgs{{\small HETGS}}
\newcommand\meg{{\small MEG}}
\newcommand\heg{{\small HEG}}
\newcommand\aciss{{\small ACIS-S}}
\newcommand\cak{{\small CAK}}

\begin{document}

\title{X-ray Spectral Study of the Photoionized Stellar Wind in Vela~X-1}
\author{Shin Watanabe\altaffilmark{1}, 
Masao Sako\altaffilmark{2}, 
Manabu Ishida\altaffilmark{3}, 
Yoshitaka Ishisaki\altaffilmark{3}, 
Steven M. Kahn\altaffilmark{2}, 
Takayoshi Kohmura\altaffilmark{4}, 
Fumiaki Nagase\altaffilmark{1}, 
Frederik Paerels\altaffilmark{5}, 
Tadayuki Takahashi\altaffilmark{1}}

\altaffiltext{1}{Japan AeroSpace Exploration Agency, 
Institute of Space and Astronautical Science, 3-1-1 Yoshinodai,
Sagamihara, Kanagawa 229-8510, Japan; watanabe@astro.isas.jaxa.jp}
\altaffiltext{2}{Kavli Institute for Particle Astrophysics and Cosmology,
 Stanford University, Stanford, CA 94305, USA}
\altaffiltext{3}{Department of Physics, Tokyo Metropolitan University, 1-1 Minami-Osawa Hachioji, Tokyo 192-0397, Japan}
\altaffiltext{4}{Kogakuin University, 2665-1 Nakano-cho, Hachioji, Tokyo 192-001
5, Japan}
\altaffiltext{5}{Columbia Astrophysics Laboratory, 550 West 120th St., New York,
 NY 10027, USA}

\shorttitle{X-ray Spectral Study of Vela X-1}
\shortauthors{Watanabe et al.}

\begin{abstract}
We present results from quantitative modeling and spectral analysis of the
high mass X-ray binary system Vela X-1 obtained with the \Chandra\ High Energy
Transmission Grating Spectrometer.  The observations cover three orbital phase
ranges within a single binary orbit.  The spectra exhibit emission lines from
H-like and He-like ions driven by photoionization, as well as fluorescent
emission lines from several elements in lower charge states.  The properties
of these X-ray lines are measured with the highest accuracy to date.  In order
to interpret and make full use of the high-quality data, we have developed a
simulator, which calculates the ionization and thermal structure of a stellar
wind photoionized by an X-ray source, and performs Monte Carlo simulations of
X-ray photons propagating through the wind.  The emergent spectra are then
computed as a function of the viewing angle accurately accounting for photon
transport in three dimensions including dynamics.  From comparisons of the
observed spectra with results from the simulator, we are able to find the
ionization structure and the geometrical distribution of material in the
stellar wind of Vela X-1 that can reproduce the observed spectral line
intensities and continuum shapes at different orbital phases remarkably well.
We find that the stellar wind profile can be represented by a \cak -model with
a star mass loss rate of (1.5--2.0)~$\times 10^{-6} M_{\sun}
\mathrm{yr}^{-1}$, assuming a terminal velocity of 1100~km~s$^{-1}$.  It is
found that a large fraction of X-ray emission lines from highly ionized ions
are formed in the region between the neutron star and the companion star.  We
also find that the fluorescent X-ray lines must be produced in at least three
distinct regions -- (1) the extended stellar wind, (2) reflection off the
stellar photosphere, and (3) in a distribution of dense material partially
covering and possibly trailing the neutron star, which may be associated with
an accretion wake.  Finally, from detailed analysis of the emission line
profiles, we demonstrate that the stellar wind dynamics is affected by X-ray
photoionization.
\end{abstract}

\section{Introduction}
\label{sec:intro}
In a high mass X-ray binary (\hmxb), a compact object, either a neutron star
or a black hole, sweeps up material as it orbits through the stellar wind of a
massive O- or B-type companion star.  When the material is accreted onto the
compact object, a fraction of its gravitational potential energy is converted
into X-rays, which in turn ionizes and heats the surrounding stellar wind.
X-ray photons from the compact object are reprocessed by the ionized material,
resulting in discrete spectral features, which carry a wealth of information
about the geometry and physical state of the material in the system.  The
compact object can, therefore, be used as a radiation source to probe the
structure of the stellar wind to derive the physical parameters that
characterize its nature.

Vela X-1 is the archetype \hmxb\ consisting of a neutron star and a massive
companion star, and as such it has been the most well-studied object in this
class since its discovery.  It is an eclipsing \hmxb\ system containing a
pulsar with a pulse period of 283~s \citep{mcclintock76} and an orbital period
of 8.964~days \citep{forman73}.  The optical companion star, HD~77581, is a
B0.5Ib supergiant \citep{brucato72, hiltner72}, which drives a stellar wind
with a mass-loss rate in the range (1 --
7)~$\times$~10$^{-6}~M_{\sun}$~yr$^{-1}$ \citep{hutchings76, dupree80,
  kallman82b, sadakane85, sato86a}.  The terminal velocity of the stellar wind
was determined to be 1100~km~s$^{-1}$ from the P-Cygni profile of UV resonance
lines \citep{prinja90}.  Its average intrinsic X-ray luminosity is
$\sim$~10$^{36}$~erg~s$^{-1}$, which is consistent with accretion of the
stellar wind by the gravitational field of a neutron star.

The first attempt to model the global X-ray emission line spectrum of Vela X-1
was presented by \cite{sako99} using the \ccd\ spectrum obtained with \asca.
They characterized the wind using a standard velocity profile of OB stars
proposed by \cite{castor75} (\cak\ model).  By adjusting the mass loss rate,
they found a statistically acceptable fit to the \asca\ spectrum.  However,
since the energy resolution of \asca\ is not sufficiently high to resolve any
of the emission lines predicted by the ionization model, the intensity
measurement of each line is limited due to blending.  In addition, since the
expected X-ray line shifts and widths are much smaller than the width of the
instrument response, they were not able to extract any information about the
dynamics of the X-ray emitting gas.

The spectral resolution ($E/\Delta E$~$\sim$~100--1000) of the transmission
grating spectrometers on board \Chandra\ have provided a detailed view of the
emission line spectrum, including accurate measurements of line intensities
and velocity profiles in a number of \hmxb s (Vela X-1:~\citealt{schulz02},
\citealt{goldstein04}; GX~301$-$2:~\citealt{watanabe03};
Cyg~X-3:~\citealt{paerels00}; 4U~1700$-$37:~\citealt{boroson03}).
\cite{schulz02} present the first high-resolution X-ray spectrum of Vela X-1
observed during eclipse.  They detect a narrow \rrc\ from \ion{Ne}{10} with an
electron temperature of $\sim$~10~eV, which implies that the stellar wind is
indeed driven by photoionization.  \cite{goldstein04} present the spectra of
Vela X-1 observed during the three orbital phase ranges -- eclipse,
$\phi$~=~0.25, and $\phi$~=~0.50.  They report the first detection of multiple
emission lines at orbital phase $\phi$~=~0.5 and analyzed the spectral changes
between the three orbital phases.

In this paper, we present a quantitative analysis of the X-ray spectra of Vela
X-1 first presented by \cite{goldstein04} using the data obtained with
\Chandra\ High Energy Transmission Grating Spectrometer (\hetgs) during three
orbital phase ranges.  We construct a simulator to compute the physical
conditions of the surrounding gas for a given binary configuration, stellar
wind parameters, and X-ray luminosity.  Continuum photons that originate from
the compact object interact with the atoms to produce emission lines and
continua, which are propagated through the wind accounting for all relevant
radiative transfer effects including line scattering/absorption, continuum
absorption, and Compton scattering.  The emerging spectra are calculated and
compared with the observed orbital-phase-resolved spectra, which provide
constraints on the global physical properties of the ionized stellar wind.

This paper is organized as follows. In \S~2, we describe the
\Chandra\ observations and summarize the data reduction procedures.  The
reduced spectra are fitted using simple empirical models to extract the
general properties of the stellar wind.  In \S~3, we provide a full
description of the simulator, which is used to model the binary system and to
calculate the resulting spectra as a function of the orbital phase.  In \S~4,
we describe a self-consistent picture of the Vela X-1 system derived from the
simulator in comparison to the data.  We finally summarize our results in
\S~5.

\section{Observations and Results}

\begin{figure}
\epsscale{0.4}
\plotone{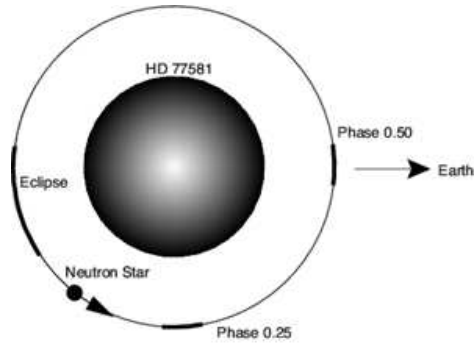}
\caption{The orbit of the neutron star in the Vela X-1 system.  The observer
is located along the horizontal axis on the right.  The bold lines span the
range of orbital phases covered by the \Chandra\ observations.}
\label{fig:velax1orbit}
\end{figure}

Vela X-1 was observed with the \Chandra\ \hetgs/\aciss\ during three different
orbital phase ranges within a single binary orbit in February 2001 -- (1)
$\phi$~=~0.237--0.278, (2) $\phi$~=~0.481--0.522, and (3)
$\phi$~=~0.980--0.093, hereafter referred to as phase 0.25, phase 0.50, and
eclipse, respectively.  The observation dates and exposure times are
summarized in Table~\ref{tbl:velax1obslist}.  A schematic drawing of the
binary orbit and the phase ranges are shown in Fig.~\ref{fig:velax1orbit}.

All of the data are processed using CIAO
v2.3\footnote{http://cxc.harvard.edu/ciao/}, and spectral analyses are
performed using
XSPEC\footnote{http://heasarc.gsfc.nasa.gov/docs/xanadu/xspec/}.  Since the
zeroth order image was severely piled-up during phase 0.25 and phase 0.50, the
locations of the zeroth order image were determined by finding the
intersection of the dispersed events as well as the streak events, which are
produced during \ccd\ frame transfer.  This gives us a zeroth order location
that is accurate to within 0.1 pixels, which is much smaller than the
instrument line response.  The positive and negative spectral orders are
inspected separately to check that systematic offsets do not exist.  We then
apply spatial filters for both the Medium Energy Grating (\meg) and the High
Energy Grating (\heg) data, and subsequently use an order sorting mask by
using the energy information obtained with \aciss.  Only the first order
events are used in our spectral analyses.  The background events are estimated
from the adjacent region to the dispersed event region and are found to
contribute at most 5~\% for the eclipse data and 3\% for the phase 0.25 and
phase 0.50 data. The background is subtracted in the spectral analyses.

\subsection{Continuum Emission}
\label{sec:continuum}

\begin{figure}
\epsscale{0.6}
\plotone{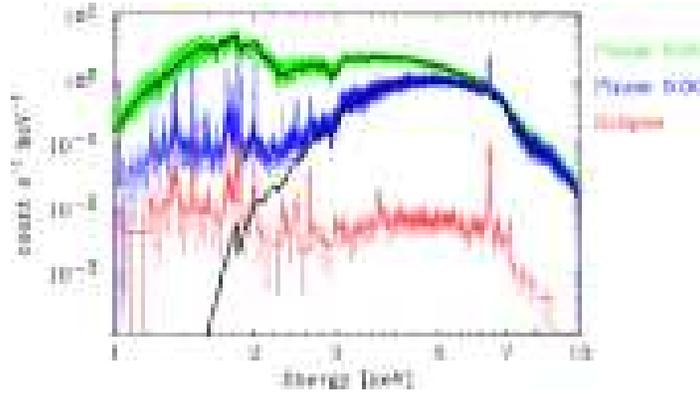}
\caption{The observed HEG spectra of Vela X-1. The green, blue, and red
histograms show the spectra at phase 0.25, phase 0.50 and eclipse,
respectively.  The solid black lines represent best-fit continuum spectral
models for phases 0.25 and 0.50 with parameters listed in
Table~\ref{tbl:cont_fit}.}
\label{fig:velacont1}
\end{figure}

Figure~\ref{fig:velacont1} shows the \heg\ spectra for the three orbital
phases.  X-rays emitted from the neutron star are directly observed during
phase 0.25 and phase 0.50 as continuum spectra.  As expected from the geometry
of the binary system, the spectrum taken in eclipse is dominated by line
emission and scattered components.  In order to parameterize the properties of
the continuum, we fit the spectra with a photo-absorbed power-law function.
Since the spectrum of phase 0.25 is affected less by absorption, we leave both
the hydrogen column density and the photon index as free parameters in the
fit.  On the other hand, for phase 0.5, we fix the photon index to the value
obtained from phase 0.25 and calculate the absorption.  We assume that the
chemical abundance of metals in the absorbing material is 0.75 times the
cosmic abundance \citep{feldman92}, which is known to be a representative
value for typical OB-stars \citep{bord76}.  The derived parameters from the
spectral fits are listed in Table~\ref{tbl:cont_fit}.  The best-fit models are
superimposed on the spectra in Fig.~\ref{fig:velacont1}.  The
absorption-corrected luminosity is determined to be identical for observations
in phase 0.25 and in phase 0.50 and corresponds to
1.6~$\times$~10$^{36}$~erg~s$^{-1}$ in the 0.5--10~keV range, assuming a
distance of 1.9~kpc \citep{sadakane85}.

\subsection{Line Emission}
\label{sec:line}
\begin{figure}
\epsscale{0.7}
\plotone{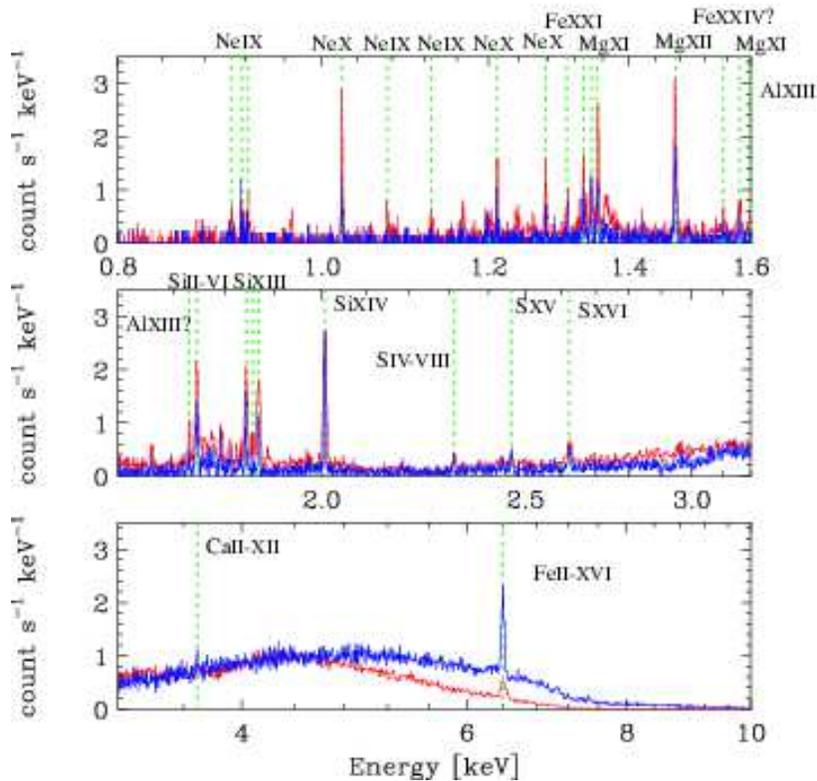}
\caption{A detailed view of the Vela X-1 spectrum at phase 0.50 -- 
MEG data in red and HEG data in blue.  The spectrum is dominated by discrete
features below $E \sim 3~\rm{keV}$.  The vertical green lines label the
observed emission lines that are listed in Table~\ref{tbl:050_line_fit}.}
\label{fig:050_line_spec}
\end{figure}

\begin{figure}
\epsscale{0.7}
\plotone{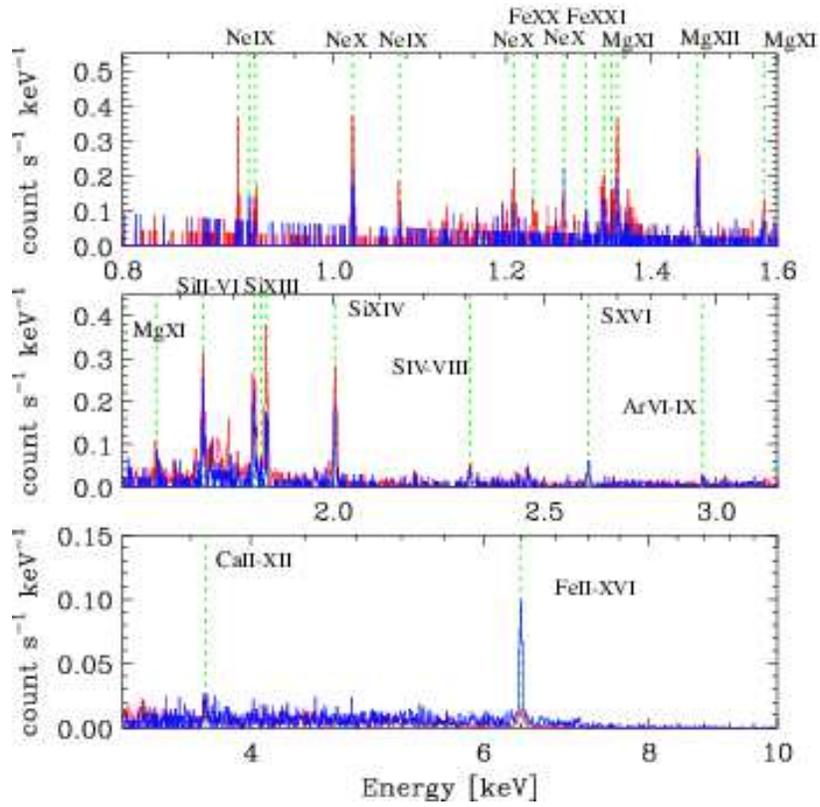}
\caption{Same as in Figure~\ref{fig:050_line_spec} for the eclipse phase.
Note the striking similarity of the global emission line spectrum compared to
that seen at phase 0.50.  The vertical green lines label the emission lines
listed in Table~\ref{tbl:000_line_fit}.}
\label{fig:000_line_spec}
\end{figure}

A large number of emission lines are clearly detected in the spectra of phase
0.50 and eclipse as shown in Fig.~\ref{fig:050_line_spec} and
Fig.~\ref{fig:000_line_spec}.  Emission lines from H-like and He-like S, Si,
Mg and Ne can be seen, in addition to fluorescent lines from Fe, Ca, S, and Si
ions in lower charge states.  We note that emission lines from the same ions
are detected in the spectra of both phase 0.50 and eclipse.

In Fig.~\ref{fig:si_line_spec}, we show the three spectra in a narrow energy
range of $E = 1.70 - 2.05$~keV, which contains the K$\alpha$ transitions of
Si.  An intense Ly$\alpha$ line from H-like Si and fully-resolved He-like
triplet lines are clearly seen in the phase 0.50 and eclipse spectra.  At
slightly lower energy, a spectacular complex of emission lines are also
detected, which we identify as K$\alpha$ fluorescent lines from Si in lower
charge states.  In the phase 0.5 spectrum, the near-neutral lines (\ion{Si}{2}
-- \ion{Si}{6}) are blended at $E \sim 1.74$~keV, and a series of lines above
this energy are resolved into those from \ion{Si}{7} -- \ion{Si}{11}.  This
implies that the stellar wind in this system contains an extremely wide range
in ionization parameter.  The eclipse spectrum also shows a similar structure,
but with much lower intensities.

The centroid of energies, the widths and the intensities of each line are
determined by representing them with single gaussian components.  In the
spectral fittings, we use the Poisson likelihood statistics, instead of the
$\chi^2$ statistics, because the numbers of photons in some of the bins in
the spectra are small.  The derived parameters for phase 0.50 and eclipse are
listed in Table~\ref{tbl:050_line_fit} and Table~\ref{tbl:000_line_fit},
respectively.  We also show list in Table~\ref{tab:comparison_lines} the line
intensity ratios between phase 0.50 and eclipse for the H- and He-like lines
that are detected with a statistical significance of more than 5~$\sigma$.
The H-like lines are brighter during phase 0.5 by a factor of 8--10 compared
to during eclipse, and similarly the He-like lines are 4--7 times stronger.

\begin{figure}
\epsscale{1.0}
\plottwo{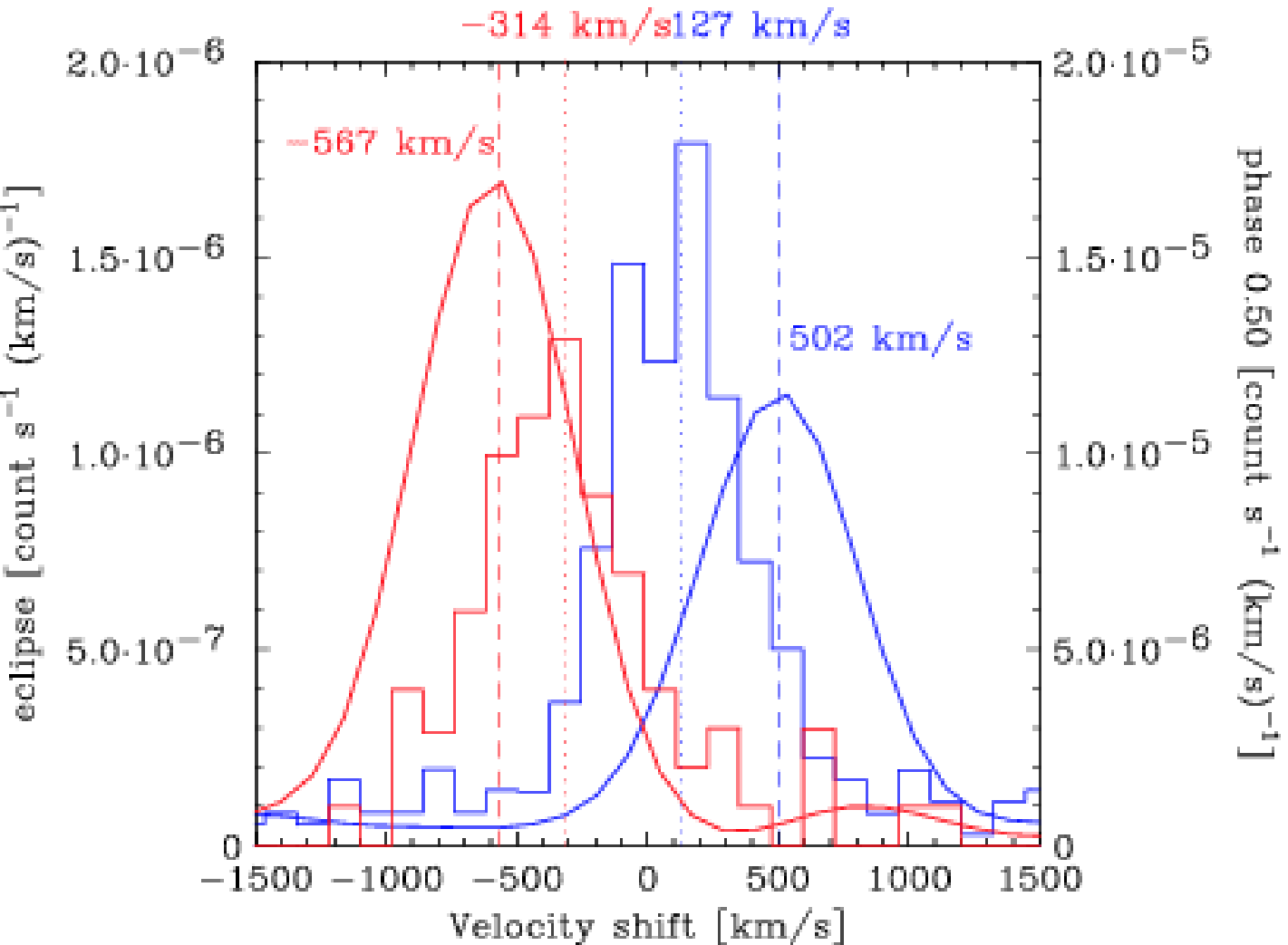}{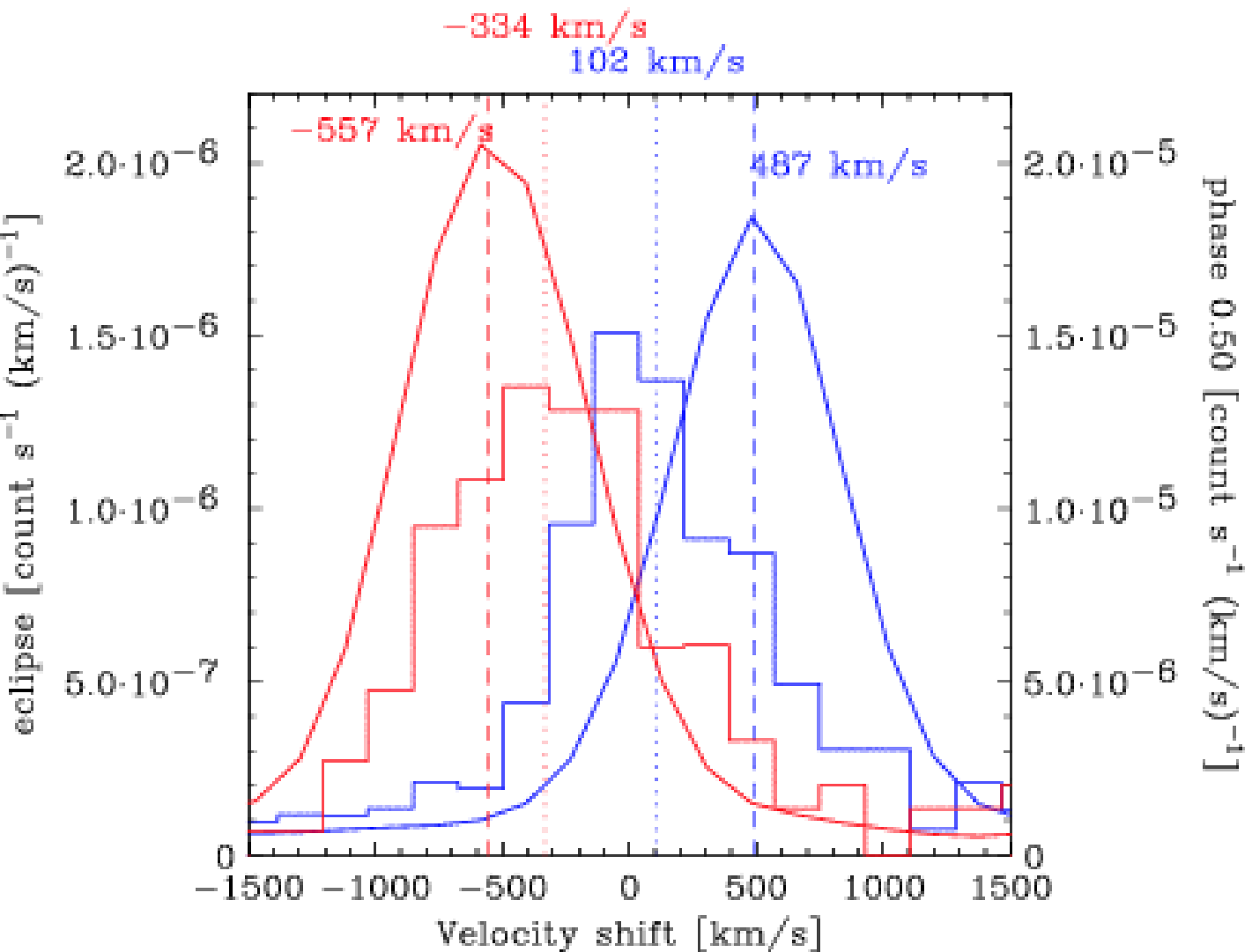}
\caption{Line profiles of Si XIV Ly$\alpha$ in the HEG spectrum (left) and
Mg XII Ly$\alpha$ in the MEG spectrum (right) plotted at a function of the
velocity.  The blue and red histograms show the observed data at phase 0.50
and eclipse, respectively, and the smooth curves were made from the output of
the Monte Carlo simulation (see \S~4.6) convolved with the respective HEG/MEG
response functions.  Note the discrepancy between the data and model on the
order of $200 - 400~\rm{km~s}^{-1}$ in the sense that the model predicts
larger velocity amplitudes than the observed values.}
\label{fig:comparison_doppler}
\end{figure}

\begin{figure}
\epsscale{0.7}
\plotone{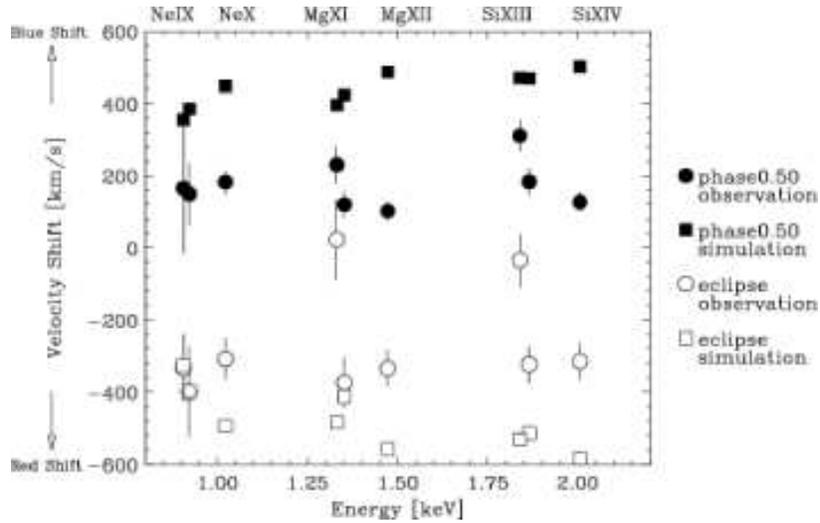}
\caption{The plot of the velocity shifts for emission lines from highly ionized
ions.  The circles represent values derived from the observations, and the
squares are values derived from the Monte Carlo simulations (see \S~4.6).
Filled and open symbols are used to designate values derived from phase 0.50
and eclipse, respectively.  As shown in Figure~\ref{fig:comparison_doppler}
for two particular lines, the lines from the other ions also systematically
show a smaller velocity amplitude compared to the simulations.}
\label{fig:dopplershift}
\end{figure}

The resolving power of the \hetgs\ allows us to measure Doppler shifts to an
accuracy of $\sim$~100~km~s$^{-1}$.  Figure~\ref{fig:comparison_doppler}
compares the line profiles of Si Ly$\alpha$ and Mg Ly$\alpha$ between phase
0.50 and eclipse.  The offset in the line center energies are clearly seen.
We measure the velocity shifts of all of the K$\alpha$ lines of H-like and
He-like ions, which are plotted in Fig.~\ref{fig:dopplershift}.  Though some
fluctuations are seen, there is a general trend that the lines are
blue-shifted during phase 0.50, and they are red-shifted during eclipse as
expected from an outflowing wind irradiated by an X-ray source.  The velocity
shifts have values in the range in $\Delta v~\sim$~300--600~km~s$^{-1}$
(Table~\ref{tab:comparison_lines} and widths of $\sigma \lesssim
300$~km~s$^{-1}$.

Radiative recombination continuum (RRC) is detected clearly from H-like Ne.
We fitted the RRC spectra using the ``redge'' model in XSPEC.  The electron
temperatures are determined to be $kT_{e} = 7.4_{-1.3}^{+1.6}$~eV and $kT_{e}
= 6.6_{-1.8}^{+2.5}$~eV during phase 0.50 and eclipse, respectively.

\begin{figure}
\epsscale{0.8}
\plotone{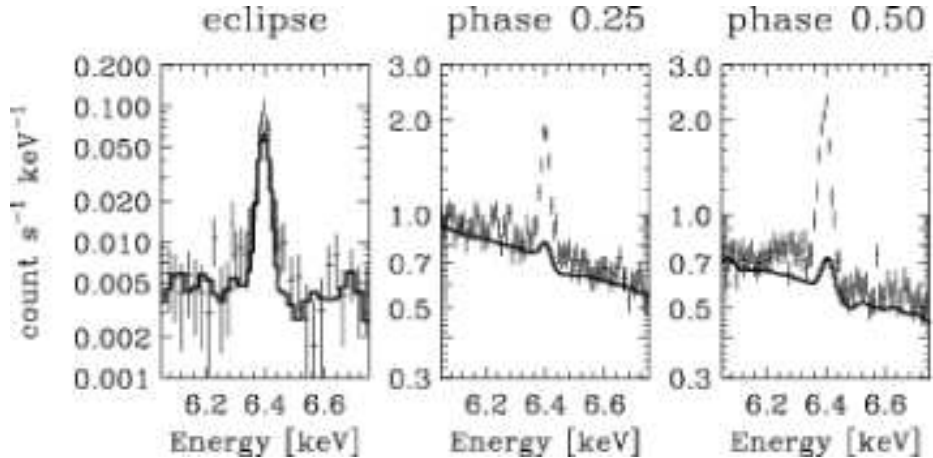}
\caption{The blown-up spectra of the iron K$\alpha$ region at the three
orbital phases.  The data points represent the HEG spectrum and the dark
histograms show the Monte Carlo models, which include iron fluorescent X-rays
from the stellar wind alone (see \S~4.5).  The model accounts for essentially
all of the line flux observed during eclipse, but fails to reproduce the
non-eclipse data by a large factor.}
\label{fig:vela_fek}
\end{figure}

Iron K$\alpha$ fluorescent lines are detected in all three orbital phases.
The profiles of these lines are shown in Fig.~\ref{fig:vela_fek}.  The
parameters derived from the spectral fits with single gaussian models are
listed in Table~\ref{tab:fekalpha}.  The equivalent width of the iron
K$\alpha$ line is measured to be 116~eV and 51~eV for phases 0.50 and 0.25,
respectively.  During eclipse, a high equivalent width of 844~eV is observed.
As shown in Fig.~\ref{fig:vela_fek}, there is an indication of a Compton
shoulder in the iron K$\alpha$ line at phase 0.50.

\subsection{Summary of Results and their Implications}
Some general global properties of the circumsource medium can be deduced from
the results of the simple fitting procedures described in
\S~\ref{sec:continuum} and \S~\ref{sec:line}.

The continuum from the neutron star is highly absorbed during phase 0.5
whereas the absorption column density is reduced during phase 0.25.  This
implies that there is an absorber located behind the neutron star as viewed
from the companion star.  The average continuum intensities are nearly
identical during the two phase ranges.  The presence of strong soft X-ray
lines during phase 0.5, however, suggests that this absorber is localized and
does not cover the X-ray emission line regions, which most likely extend
beyond the size of the absorber.

The emission lines during phase 0.50 is approximately an order of magnitude
brighter than those during eclipse.  This indicates that a significant
fraction of the X-ray line emission is produced in a region between the
neutron star and the companion star, which is occulted during eclipse.  The
velocity shifts and widths provide additional constraints on the properties of
the emission region.  The measured Doppler shifts of $\la 500$~km~s$^{-1}$ are
much smaller than the terminal velocity of Vela X-1 ($v_\infty =
1100$~km~s$^{-1}$; \citealt{prinja90}).  For a \cite{castor75} stellar wind
velocity profile of isolated OB stars, this velocity is achieved at only 0.25
stellar radii from the photosphere.  The Doppler broadening of emission lines
is also small ($\la 300$~km~s$^{-1}$).  These observational results shows that
the emission site of lines from highly ionized ions is distributed on a plane
between the neutron star and the companion star, with most of the emission
coming from near the line of centers.

Finally, the detection of a narrow radiative recombination continuum from
H-like Ne during phase 0.5 and in eclipse with electron temperatures of
6--8~eV provides direct evidence that the emission lines from highly ionized
gas in Vela X-1 are driven through photoionization.  Other possible sources of
mechanical heating are, therefore, not important in characterizing the
ionization structure of at least the highly-ionized regions of the stellar
wind.

\section{Outline of the Simulator}
\label{sec:simulator}
\subsection{Modeling of Photoionized Plasma}

The X-ray emission lines detected from Vela X-1 can be interpreted as due to
emission from a gas photoionized by the continuum radiation from the neutron
star.  When line photons are created through recombination cascades following
photoionization, they interact with the surrounding gas as they propagate out
of the binary system, and this is controlled by the ionization structure and
the density distribution of the gas in the stellar wind.  Therefore, by
investigating the properties of the emission lines, we can infer the physical
conditions, geometrical distribution, and the velocity structure of the wind.

In order to compute the ionization structure and to reproduce the wealth of
information provided by the high-resolution \hetgs\ spectra, it is necessary
to model the emission using realistic physical models of the stellar wind in a
binary system.  We have developed
three-dimensional Monte Carlo radiative transfer code to simulate the
interactions of X-rays with the particles in the stellar wind.  

Our simulation code consists of two main components:
\begin{description}
  \item[(1)] Calculation of the ionization structure.
  \item[(2)] Monte Carlo simulation for radiative transfer.
\end{description}
In part (1), we construct a map of the ion abundance and temperature in the
stellar wind.  In part (2), we use the results from part (1) and propagate
X-ray photons from the neutron star out of the binary system.  The simulation
is performed on a three-dimensional grid of cells, which is capable of
treating complex geometries.  Radiative transfer effects that are important
for highly ionized media are included.  The medium is allowed to have
arbitrary velocities, and Lorentz transformations are performed between the
cells, which is necessary to calculate the correct output line profiles.  Our
detailed procedures are described in the following subsections.

\subsection{Calculation of the Distribution of Ionization Degree 
\label{sec:calcionization}}

\begin{figure}
\epsscale{0.5}
\plotone{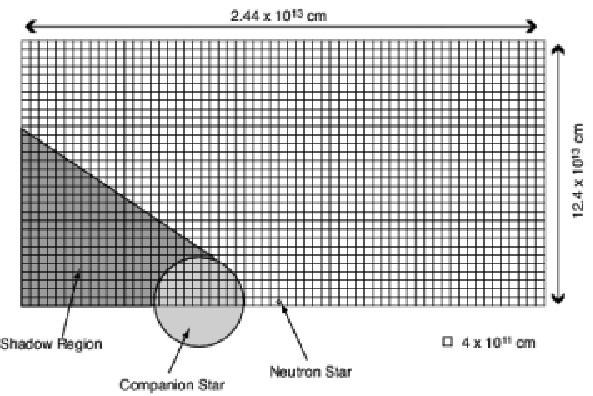}
\epsscale{0.3}
\plotone{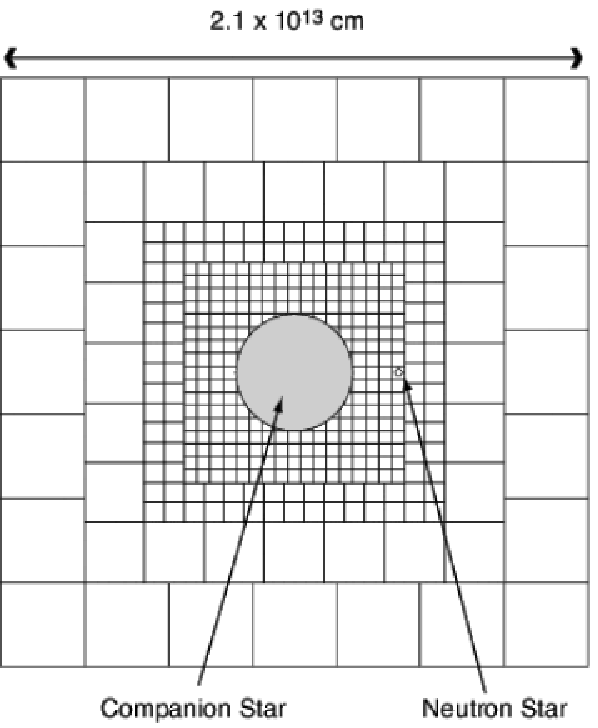}
\caption{The Vela X-1 geometry and spatial grids used in the calculation of
the ionization structure (left) and the Monte Carlo spectral simulation
(right).}
\label{fig:geom_grid}
\end{figure}

The ionization balance and electron temperature are determined primarily by
the particle density and the local flux and spectral shape of the ionizing
X-rays.  We compute the three-dimensional ionization structure by first
dividing the binary system into square cells of equal size.  Since the system
is assumed to be cylindrically symmetric along the line of centers, we compute
the ion fractions and temperatures on a two-dimensional grid as shown in
Fig.~\ref{fig:geom_grid}.  Each cell is $4 \times 10^{11}$~cm along the side
and the system consists of $61 \times 31$ cells. 

\begin{figure}
\epsscale{0.7}
\plotone{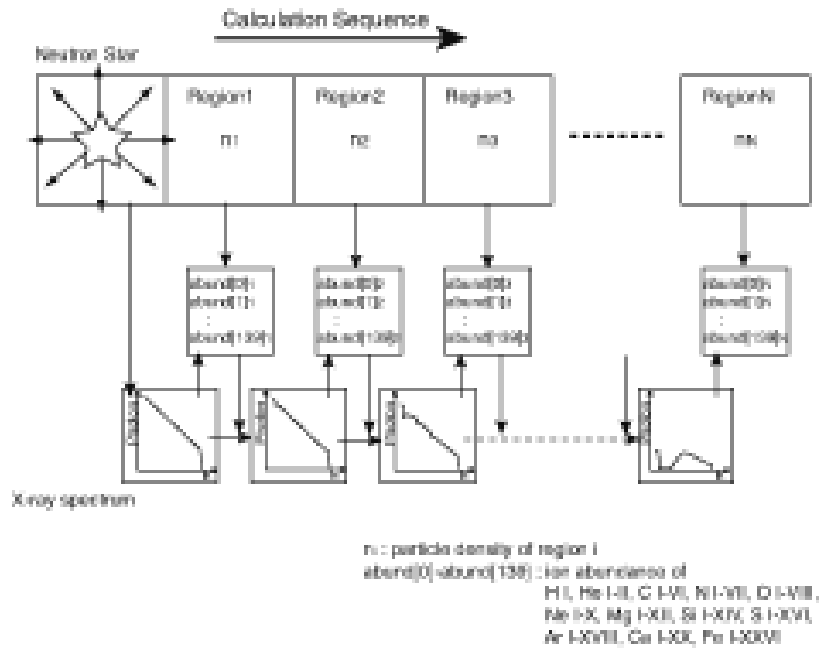}
\caption{A schematic diagram showing the sequence of how the
ionization structure is calculated in the Monte Carlo simulator.  The
continuum photons from the neutron star (left) is propagated through each
spatial cell and transfered to the next cell accounting for absorption.}
\label{fig:sim_scheme1}
\end{figure}

The density structure around the companion star is specified from the velocity
profile determined from the UV observation and an assumed mass-loss rate,
which is treated as a free parameter.  This assigns a particle density to each
of the cells.  The intrinsic luminosity and spectrum shape of the X-ray source
as determined from the observed spectrum are then placed in the cell where the
neutron star is located.  The ion fractions and the electron temperatures of
the cells immediately surrounding the neutron star are computed using this
initial continuum shape as the input ionizing radiation field.  The
computation moves outwards to the next layer of cells, this time, using a
different continuum shape, which accounts for absorption by the inner cells.
This procedure continues until the outermost cells are reached.  We have use
the
XSTAR\footnote{http://heasarc.gsfc.nasa.gov/docs/software/xstar/xstar.html}
program to calculate the ion fractions H, He, C, N, O, Ne, Mg, Si, S, Ar, Ca,
and Fe, and the electron temperature in each cell.  A flowchart of the
procedures is shown in Fig.~\ref{fig:sim_scheme1}.

This procedure ignores the diffuse radiation field as the source of additional
ionization, which is a small contributor to the total photoionization rate at
least for the ions observed in our spectra.  A more accurate treatment of this
effect might involve an iterative scheme, which is not necessary for our
present analysis, but may be important for extremely optically thick media.

\subsection{Monte Carlo Calculation of the X-ray Emission from 
Photoionization Equilibrium State}

The map of the ion fractions calculated by the method described in the
previous subsection assures equilibrium between photoionization and
recombination.  We then use this ionization structure to propagate again the
continuum photons originating from the neutron star to compute the emitted
X-ray spectra as a function of the orbital phase.  This is done by accounting
for radiative transfer effects in detail and using Monte Carlo methods as
described below.

The cells are set up somewhat differently from those used in the calculation
of the ionization structure.  The cells sizes are chosen so that each of them
are not optically thick to the strongest resonance lines.  Smaller cells are
required near the companion star where both the density and density gradient
are high, and we choose cubic cell sizes of $4 \times 10^{11}$~cm on each
side.  The cells are increased in size outwards and the outermost cells of the
simulation space are larger by a factor of 7.5 compared to the inner ones as
shown in Fig.~\ref{fig:geom_grid}.  We start the Monte Carlo simulation with a
photon at the position of the central neutron star.  The photon energy is
drawn from the observed power-law spectral energy distribution and is
initially given a random direction.  This photon is then traced through the
cells and interacts with the ions in the stellar wind, which then generates a
secondary photon through radiative recombination, radiative transitions, and
fluorescent emission.  All of the photons including those from the central
continuum and subsequent line photons produced by interactions are propagated
until they either completely escape the simulation space or are destroyed by
one of the physical processes described below.  The emergent photons are then
selected depending on the viewing angle of the system, and are finally
histogrammed to produce a spectrum.

When a photon interacts with a particle, one of three physical processes is
allowed to occur: (1) photoionization, (2) photoexcitation, and (3) Compton
scattering.  For a given photon energy, the absorption and scattering cross
sections and hence the optical depths are calculated for each ion within a
given cell.  Interaction probabilities are then assigned to each possible
process and are drawn via Monte Carlo.  Although process (1) produces a free
electron and process (3) changes the energy of a free electron, we do not
trace them in our simulations, but instead simply assume that they are
thermalized before any further interaction takes place.

As mentioned earlier, we assume that photoionization equilibrium is
established locally everywhere in the plasma.  Therefore, photoionization of
H- or He-like ions is always followed by radiative recombination and radiative
cascades to the ground level.  For ions of lower charge states, K-shell
photoionization is followed by either fluorescent emission or Auger decay.
After photoexcitation, the ion in an excited state produces one or more
photons in the downward radiative transitions that eventually lead to the
ground level. By the combination of 
the photoexcitation and the downward radiative transitions, 
the resonance scattering of the lines is incorporated in the Monte
Carlo simulation.

For the H-like and He-like ions, we compute all of the atomic quantities with
the Flexible Atomic Code ({\small
  FAC})\footnote{http://kipac-tree.stanford.edu/fac/}.  This includes energy
levels, oscillator strengths, radiative decay rates, photoionization cross
sections from all levels between the ground state up to $n = 10$.  Radiative
recombination rates onto each of levels as a function of the electron
temperature are also computed from the photoionization cross sections using
the Milne relation.  Collisional transfers and UV photoexcitations in the
He-like ions are not included in the current version of the code.

Photoionization of ions with three or more electrons de-excites by either an
emission of fluorescent photon or ejection of one or more Auger electrons.
The K-shell photoionization cross sections are computed using the fitting
formulae provided by \cite{band90}, which roughly accounts for the change in
the K-edge energy and cross section as functions of the charge state.  The
fluorescence yields and K$\alpha$ energies, however, are assumed to be fixed
for a given element and their values are adopted from \citep{larkins77} and
\citep{salem74, krause79}, respectively, for neutral atoms.  We also do not
account for K-shell ionization of Li-like ions, which can lead to the
production of the He-like forbidden line.  We use L-shell cross sections from
{\small EPDL}97
\footnote{http://www.llnl.gov/cullen1/photon.htm, and distributed with Geant4
  low energy electromagnetic package}, which is also valid for neutral atoms.
The L$_{1}$, L$_{2}$ and L$_{3}$-shell cross sections are taken into account
for the ions with three, five, and six electrons, respectively.  Finally, we
do not include L-shell fluorescent emission, since the yields are typically
much smaller compared to the K-shell fluorescence yields.

We consider Compton scattering only by free electrons, which is an excellent
approximation for highly ionized media.  The differential cross section is
given by the the Klein-Nishina formula and the electrons velocities are drawn
from a Maxwellian distribution with the local electron temperature.

Doppler shifts due to the motion of the stellar wind are accounted for in all
of the processes.  When a photon enters a cell, the cross sections are
calculated for the photon energy in the co-moving frame and Lorentz
transformed back into the rest frame. The Doppler shifts could 
affect the line strength through the resonance line scattering.
A velocity gradient can change the ratio of the line escape. 
In this Monte Carlo simulation, such Doppler effects on the lines are 
included, though the reproducibility of the velocity gradient is 
limited by the geometry cell size.

\section{The Physical Properties of the Stellar Wind in Vela X-1}
\subsection{Structure of the Stellar Wind}
We first search for the value of the mass loss rate that reproduces the
observed line intensities as a function of the orbital phase.  We assume that
the velocity structure of the stellar wind is represented by a generalized
\cak\ \citep{castor75} profile:
\begin{equation}
\label{eq:stellarwindv}
  v(r) = v_{\infty}\left( 1 - \frac{R_{*}}{r} \right)^{\beta},
\end{equation}
where $r$ is the distance from the center of the star, $R_{*}$ is the stellar
radius, and $v_{\infty}$ is the terminal velocity.  We use a terminal velocity
of 1100~km~s$^{-1}$ as determined from UV observations of Vela X-1
\citep{prinja90} and fix the value of $\beta$ to 0.80 \citep{pauldrach86}.

For a given velocity profile and a mass loss rate $\dot{M}_{*}$, the wind
density can be calculated by applying the equation of mass continuity assuming
spherical symmetry,
\begin{equation}
  n(r) = \frac{\dot{M}_{*}}{4\pi \mu m_{p} v(r) r^{2}},
\end{equation}
where $\mu$ is the gas mass per hydrogen atom, which is $\sim 1.3$ for cosmic
chemical abundances.  This specifies the density at each point in the stellar
wind, and the observed X-ray luminosity can be used to obtain a map of the
ionization parameter.

The binary system parameters used in the simulations are listed in
Table~\ref{tbl:simparams}.  The X-ray radiation field from the neutron star is
modeled based on the parameter determined from the fits to the observed
spectra at phase 0.25 and phase 0.5.  Since no change in the average intrinsic
luminosity is observed between these phases, we assume that the luminosity
remains unchanged during the eclipse phase as well.  The power-law spectrum
with a photon index $\Gamma = 1$ is assumed to extend up to 20~keV, which is
the cut-off energy detected by previous hard X-ray observations with {\it
  Ginga} \citep{makishima99} and {\small {\it RXTE}} \citep{kreykenbohm99}.
The corresponding 0.5--20~keV luminosity is
3.5~$\times$~10$^{36}$~erg~s$^{-1}$.  In the following subsection, we discuss
the results of our simulation for mass loss rates of 5.0~$\times$~10$^{-7}$,
1.0~$\times$~10$^{-6}$, 1.5~$\times$~10$^{-6}$, and
2.0~$\times$~10$^{-6}$~$M_{\sun}$~yr$^{-1}$.

\subsection{The Ionization Structure \label{sec:velaion}}

\begin{figure}
\epsscale{1.0}
\plotone{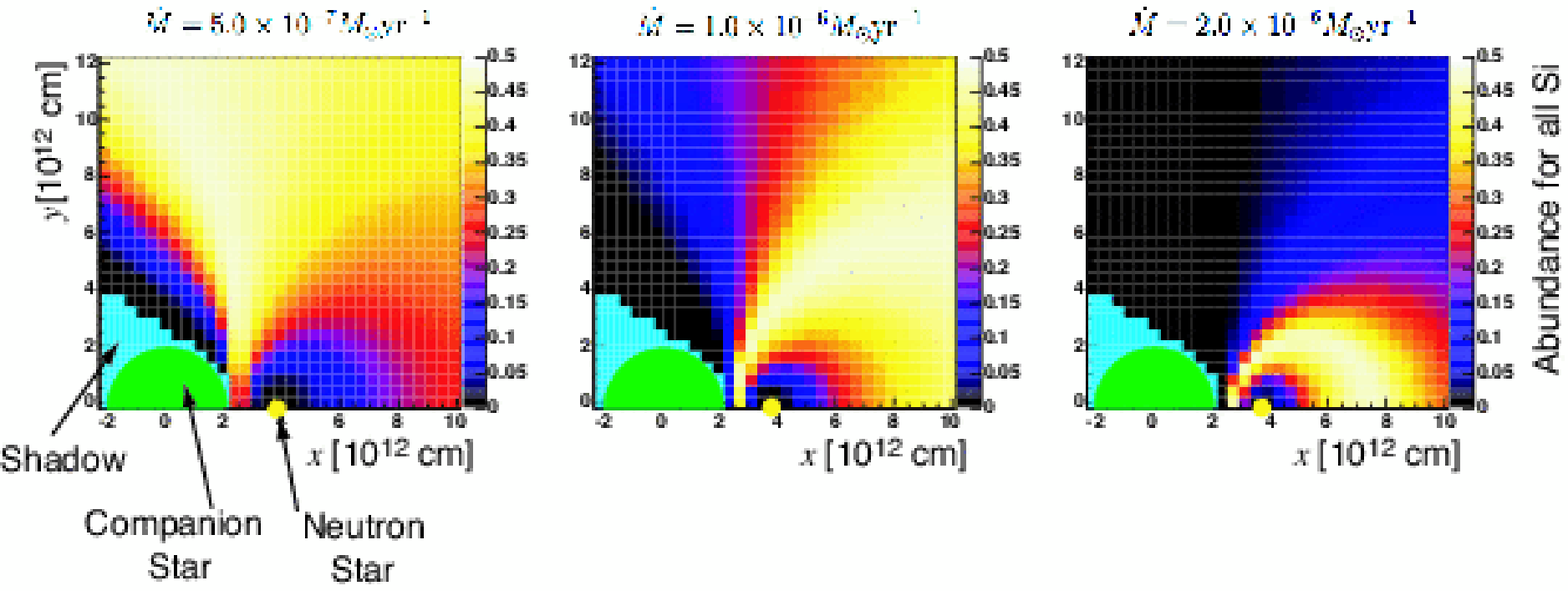}
\caption{The map of the H-like ion fraction of Si.}
\label{fig:map_abund}

\plotone{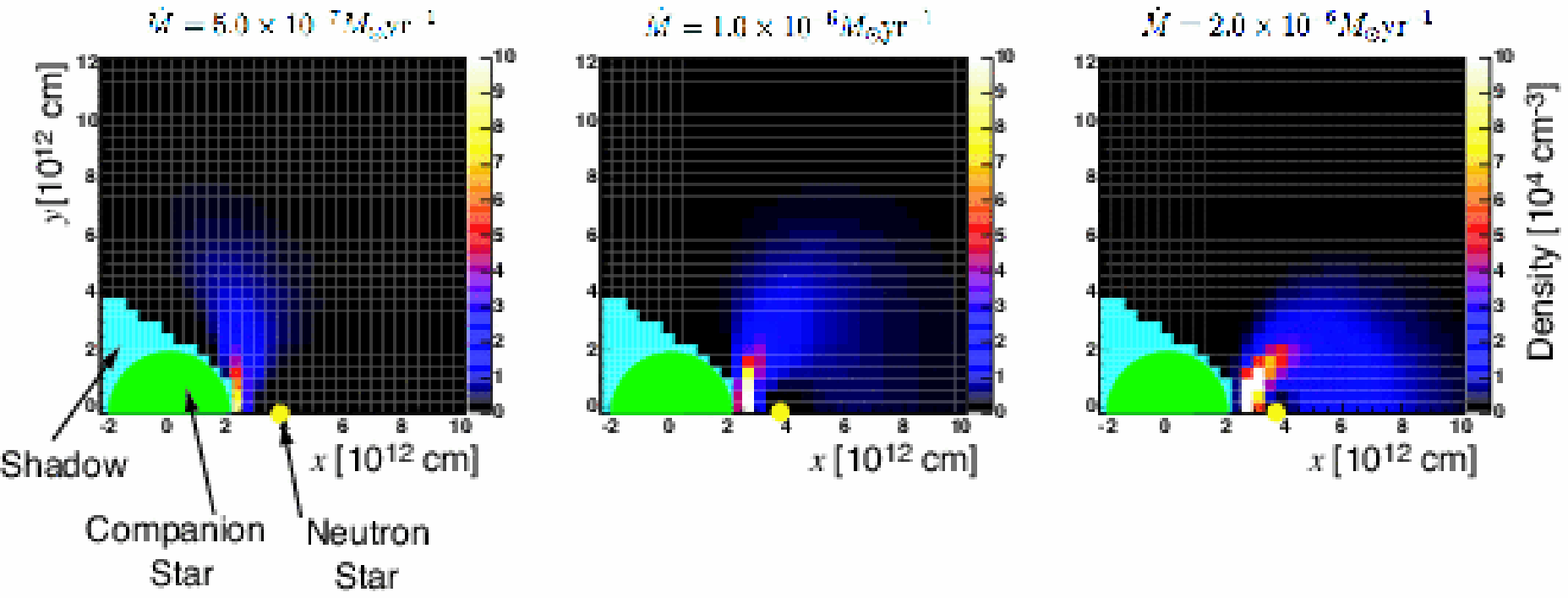}
\caption{The density map of H-like Si.}
\label{fig:map_density}

\end{figure}

Figure~\ref{fig:map_abund} shows maps of the H-like ion fraction of Si
$f_{\mathrm{Si H-like}}$ at three different values for the mass loss rate.  At
a mass loss rate of 5.0~$\times$~10$^{-7}$~$M_{\sun}$~yr$^{-1}$, the
concentration of H-like Si is highest near the mid-plane of the binary system
as shown on the left panel of Fig.~\ref{fig:map_abund}.  As the mass loss rate
is increased, the density everywhere in the wind is also increased.  For a
fixed ionization parameter, the distance to the neutron star has to be
reduced, so the region moves closer towards the neutron star as shown in the
middle and right panels of Fig.~\ref{fig:map_abund}. 
Multiplying the maps in Fig.~\ref{fig:map_abund} by the local density of the
stellar winds gives the maps in Fig.~\ref{fig:map_density}. These show the 
absolute abundances of H-like Si $n_{\mathrm{Si H-like}}$ and the region where
the X-ray photons of the emission line are produced.

According to these simulations, a large fraction of emission lines from highly
ionized ions such as H-like Si are produced in the region between the
companion star and the neutron star.  This is the region that is obscured by
the companion star during eclipse.  It turns out that the mass loss rate is 
sensitive to the ratio of line fluxes between phase 0.5 and eclipse,
because the size of the region becomes larger when we increase the mass loss
rate.  If a different value for the terminal velocity is assumed, the mass
loss rate will differ such that the quantity $\dot{M}_{*}/v_{\infty}$ remains
fixed.

\subsection{Estimate of the Mass Loss Rate}

\begin{figure}
\epsscale{0.4}
\plotone{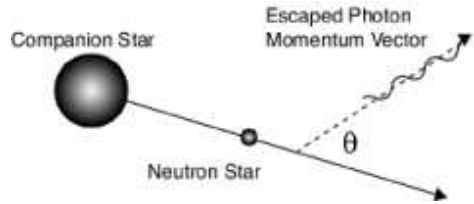}
\caption{The definition of the angle $\theta$, the angle of the outgoing
photon relative to the line of centers.}
\label{fig:direction}
\end{figure}

The simulations are run for the various values of the mass loss rate.  In
order to calculate the observed spectra at different orbital phases, we
designate the direction of the outgoing photon by the angle $\theta$ that the
momentum vector makes relative to the line of centers (see,
Fig.~\ref{fig:direction}).  The phase 0.5 spectrum is constructed from the
outgoing photons with $\theta < 10\arcdeg$ and the eclipse spectrum is made
from photons with $170\arcdeg < \theta < 180\arcdeg$.  In this calculation,
photons directly coming from the neutron star are neglected since only the
emission lines and scattered components are of interest.

Although many emission lines are resolved in the Chandra spectra, we use only
the \ion{Si}{14} Ly$\alpha$ line to estimate the mass loss rate.  This line is
one of the brightest lines observed in the data.  Since the energy of this
line is sufficiently high, the uncertainties associated with foreground
absorption is lower than the low energy lines.  At this energy, an uncertainty
of 1~$\times$~10$^{21}$~cm$^{-2}$ in $N_\mathrm{H}$, corresponds to only a
$\sim 5$\% uncertainty in the Ly$\alpha$ intensity.

\begin{figure}
\epsscale{0.5}
\plotone{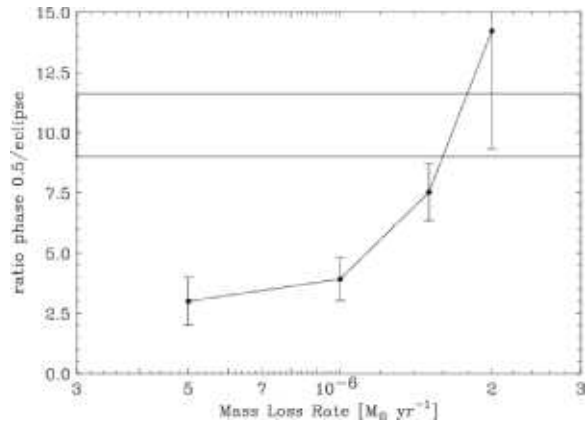}
\caption{Intensity ratio of Si Ly$\alpha$ between phase 0.5 and
eclipse as a function of the mass-loss rate.  Each point represents results
derived from the Monte Carlo simulations, and the error bars corresponds to
1$\sigma$ Poisson error due to the limited number of trials.  The two
horizontal lines denote the 1$\sigma$ range of the observed value.  This
figure shows that the mass-loss rate in the Vela X-1 system is $\sim 1.5 -
2.0 \times 10^{-6}~M_\odot~\rm{yr}^{-1}$.}
\label{fig:sih_05_ec}
\end{figure}

\begin{figure}
\epsscale{0.8}
\plottwo{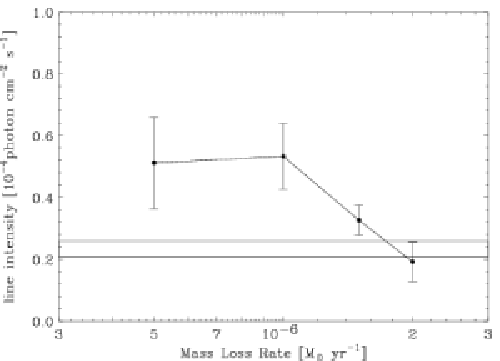}{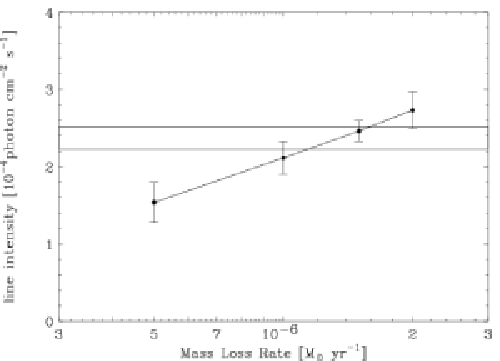}
\caption{The Si Ly$\alpha$ line intensity during
eclipse (left) and for phase 0.50 (right) as a function of the mass-loss rate.
The horizontal lines denote the 1$\sigma$ ranges of the observed values.  As
in Figure~\ref{fig:sih_05_ec}, a consistent solution can be obtained for a
mass-loss rate of $\sim 1.5 - 2.0 \times 10^{-6}~M_\odot~\rm{yr}^{-1}$.}
\label{fig:sih_line_intensity}
\end{figure}

The simulated intensity ratio of the \ion{Si}{14} Ly$\alpha$ lin between phase
0.5 and eclipse as a function of the mass loss rate is shown in
Fig.~\ref{fig:sih_05_ec}.  The errors associated with these points are due to
the finite number of photons generated in the simulation.  The horizontal
lines indicate the $1 \sigma$ range in the observed value.  As mentioned
earlier in \S~\ref{sec:velaion}, the ratio becomes larger as the mass loss
rate is increased.  The intensities of the Si Ly$\alpha$ in the phase 0.5 and
in the eclipse are shown in Fig.~\ref{fig:sih_line_intensity}. 
From these three plots
in Fig.~\ref{fig:sih_05_ec} and Fig.~\ref{fig:sih_line_intensity}, the mass
loss rate is estimated to be (1.5--2.0)~$\times 10^{-6} M_{\sun}
\mathrm{yr}^{-1}$.

The estimated mass loss rate of the stellar wind
is consistent with the observed X-ray continuum luminosity.  The mass
accreting rate onto the neutron star $\dot{M}_{acc}$ is given by
\begin{equation}
\label{eq:mdotacc}
\dot{M}_{acc} = \frac{\dot{M}_{*}R_{acc}^{2}}{4D^{2}},
\end{equation}
where $\dot{M}_{*}$ is the mass loss rate of the stellar wind, and $D$ is the
distance of the neutron star from the center of the OB star. $R_{acc}$ is the
accretion radius express as:
\begin{equation}
\label{eq:Racc}
R_{acc} = \frac{2GM_{ns}}{v_{rel}^{2}},
\end{equation}
where $v_{rel}$ is the relative velocity of the neutron star and the stellar
wind.  The gravitational energy of the accreting material is converted into
X-rays. The X-ray luminosity resulting from this accretion will simply be the
rate at which gravitational energy is released:
\begin{equation}
\label{eq:Lx}
L_{x} = 
\frac{GM_{ns}\dot{M}_{acc}}{R_{ns}}=\frac{\left(GM_{ns}\right)^{3}\dot{M}_{*}}
{R_{ns}v_{rel}^{4}D^{2}}
\end{equation}
where it is assumed that most of this energy is liberated near the neutron
star surface (of radius $R_{ns}$).  If $M_{ns} = 1.7 M_{\sun}$
\citep{vanparadijs77, barziv01}, $R_{ns} = 10$~km, $v_{rel} = 640$~km~s$^{-1}$
($v_{wind} = 570$~km~s$^{-1}$, $v_{orbit} = 300$~km~s$^{-1}$), $D = 53.4
R_{\sun}$ and $\dot{M}_{*} = 1.5 \times 10^{-6} M_{\sun} \mathrm{yr}^{-1}$ are
applied, an X-ray luminosity of $L_{X} = 4.7 \times 10^{36}$~erg~s$^{-1}$ is
obtained, which agrees with the observed luminosity of
3.5~$\times$~10$^{36}$~erg~s$^{-1}$ within a factor of 2.

From the eclipse spectrum obtained by \asca, \cite{sako99} estimated that the
mass loss rate associated with the highly ionized component of the stellar wind is
$2.7 \times 10^{-7} M_{\sun} \mathrm{yr}^{-1}$, which is a factor of 5--8
lower than the current estimate.  The lower mass loss rate was derived most
likely because only the eclipse data was used.
As the mass loss rate is increased, the intensities of the emission lines
from highly ionized ions become higher. When the mass loss rate
is increased further, the photoionized region is occulted by the companion
star and the intensities of the emission lines during the eclipse phase 
become lower. This tendency for the \ion{Si}{14} is shown 
in Fig.~\ref{fig:sih_line_intensity} (left). 
There were two solutions for the eclipse spectrum, and,
\cite{sako99} found the solution of the lower mass rate.
However, we note that the
intensity of emission lines in the phase 0.50 obtained with \Chandra\ cannot
be explained with such a lower mass loss rate. We, therefore, have concluded
the higher mass loss rate of 1.5--2.0~$\times 10^{-6} M_{\sun}\mathrm{yr}^{-1}$.
These measurements were probably
not possible with the moderate spectral resolution of the instruments on board
\asca.

\subsection{The Global X-ray Spectrum \label{sec:vela_mcsim}}

Adopting a mass loss rate of 1.5~$\times$~10$^{-6}$~$M_{\sun}$~yr$^{-1}$
estimated from the \ion{Si}{14} line, we use the results of the Monte Carlo
simulation to generate broadband spectra including both the continuum and
emission lines from other ionic species for the three orbital phases observed
with the \hetgs.  Photons with outgoing momenta in the $\theta$ ranges of
0\arcdeg--10\arcdeg, 85\arcdeg--95\arcdeg, and 170\arcdeg--180\arcdeg\ are
used for the phase 0.50, phase 0.25, and eclipse model spectra, respectively.

\begin{figure}
\epsscale{0.6}
\plotone{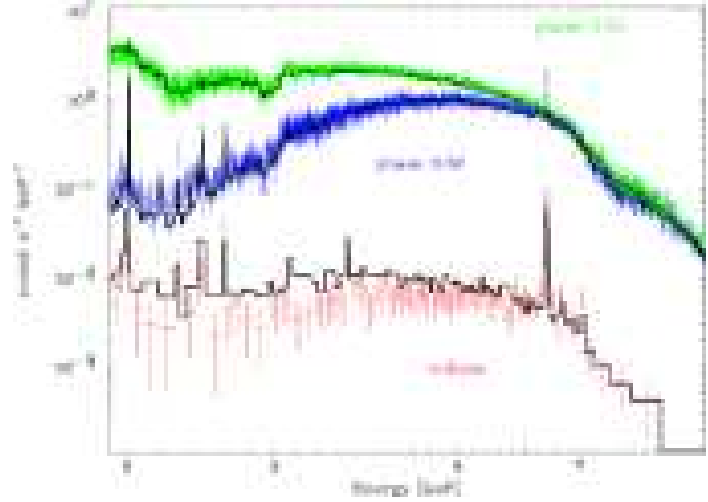}
\caption{Comparison of the simulated spectra with the observed data above
2~keV.  The normalization of the simulation models are fixed between the three
phases ranges, assuming that the average X-ray luminosity does not change.
The simulated models are multiplied by a Galactic column density of
$N_{\mathrm{H}}$~=~6~$\times$~$10^{21}$~cm$^{-2}$ and convolved with the
instrument response.}
\label{fig:all_data_model}
\end{figure}

Figure~\ref{fig:all_data_model} shows both the observed spectra above $\sim
2$~keV and the simulated model spectra convolved through the response of the
\hetgs .  The simulated spectra are multiplied by the effects of inter-stellar
absorption ($N_{\mathrm{H}} = 6 \times 10^{21}$~cm$^{-2}$) in addition to
intrinsic absorption by the stellar wind, which depends on the observed binary
phase.  This matches the observed continuum shape remarkably well at both
phases 0.25 and eclipse.  At phase 0.5, however, the observed column density
of $N_{\mathrm{H}} = 1.85 \times 10^{23}$~cm$^{-2}$ is in excess of the
Galactic plus stellar wind by an amount $N_{\mathrm{H}} = 1.7 \times
10^{23}$~cm$^{-2}$.  We apply this additional absorption component to the
continuum emission to match the data.

\begin{figure}
\epsscale{0.9}
\plotone{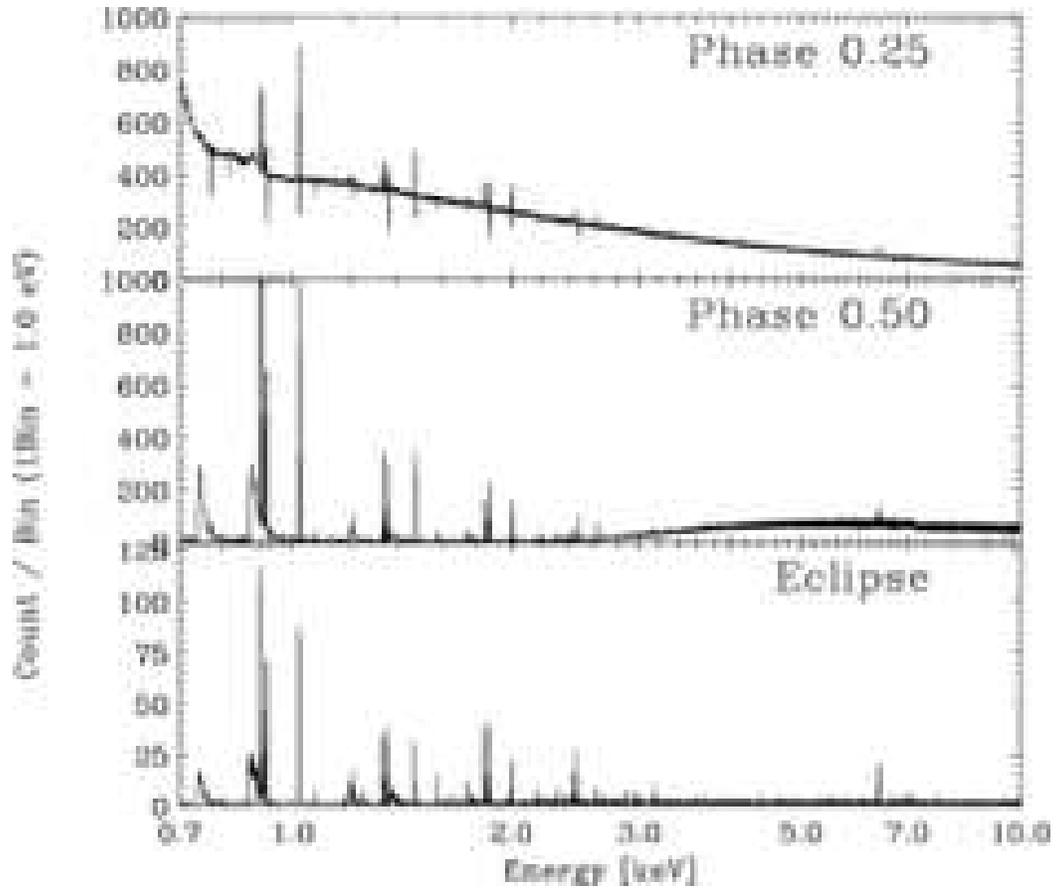}
\caption{Monte Carlo simulation spectra at the three phases 
for the mass loss rate of 1.5~$\times$~10$^{-6}$~$M_{\sun}$~yr$^{-1}$.
These spectra are not convolved with the instrumental response.}
\label{fig:sim_spec}
\end{figure}

\begin{figure}
\epsscale{1.0}
\plotone{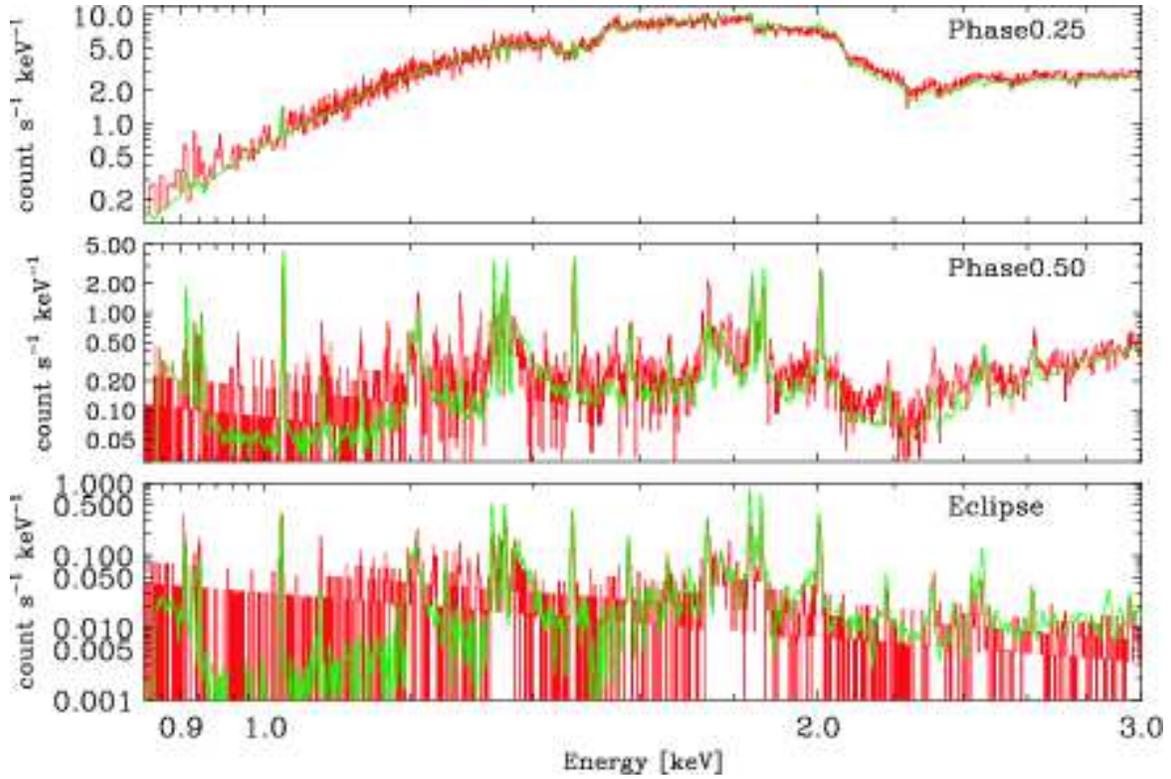}
\caption{Emission line spectra obtained with Chandra MEG. The red curves represent the data and the green lines show the
Monte Carlo data convolved with MEG instrument response.}
\label{fig:simobs_spec}
\end{figure}

\begin{figure}
\epsscale{0.6}
\plotone{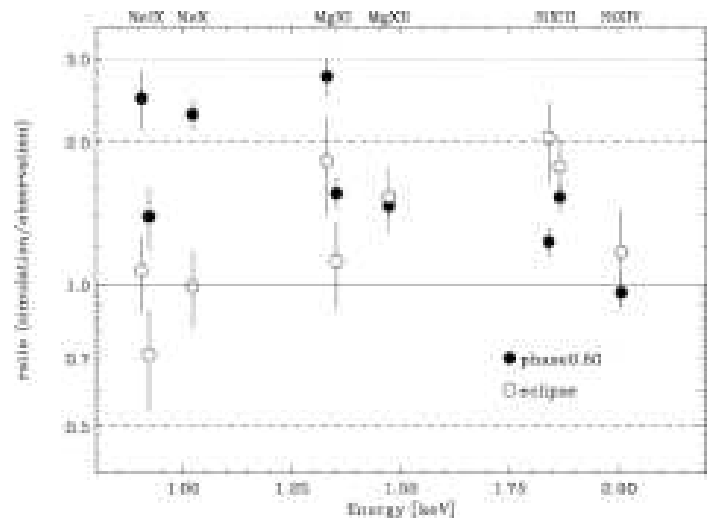}
\caption{The plots of the intensity ratios of the simulated lines and the
observed lines.  The filled and open circles show the results from phase 0.50
and eclipse, respectively.  Each of the measured lines intensities are within
a factor of 3 from the predicted values. }
\label{fig:compare_intensity}
\end{figure}

The simulated spectra at the three orbital phases are shown in
Fig.~\ref{fig:sim_spec}.  The observed emission line spectra in comparison
with the simulations are shown in Fig.~\ref{fig:simobs_spec}.  
The lack of obvious discrete spectral features at phase 0.25 is
consistent with the model predictions, which show P-Cygni lines with
comparable emission and absorption equivalent widths and are therefore washed
out after convolution with the spectral response.  We note that the additional
absorption component observed at phase 0.50 does not appear to be affecting
the emission lines, which implies that this absorber is localized to a small
region along the line of centers and in front of the neutron star at phase
0.5.  The simulated spectra match fairly well the emission lines from H-like
and He-like ions, as well as the radiative recombination continuum emission.
The ratios of simulated line intensities to the observed ones are plotted for
those lines that are detected at greater than 5~$\sigma$ in
Fig.~\ref{fig:compare_intensity}.  All of the simulated line intensities are
within the observed values to within a factor of $\sim 3$. One of the 
uncertainties of the line intensities is the absorption, and it is difficult
to derive the line intensities precisely by unfolding the absorption effect.
In the simulation, the effects of inter-stellar absorption 
($N_{\mathrm{H}} = 6 \times 10^{21}$~cm$^{-2}$)
and intrinsic absorption by the stellar wind 
($N_{\mathrm{H}} \sim 1.5 \times 10^{22}$~cm$^{-2}$) are assumed. 
If there is an additional absorption of 10\% of the assumed value, 
the intensities of Ne lines are reduced to $\sim$~50\%.
Additionally, the absorption cross sections are very sensitive to 
the ionization degree of the absorption matter.

\subsection{The Iron K$\alpha$ Line Complex}

Fluorescence emission lines, in particular the bright iron line complex,
provide information about the distribution of cold material around the X-ray
source.  Figure~\ref{fig:vela_fek} shows the iron K$\alpha$ line spectrum at
the three orbital phases, together with the models obtained from the
simulations described in the previous subsection.  During eclipse, the iron
K$\alpha$ line intensity calculated from the simulation is consistent with the
observed one.  Therefore, it seems likely that the iron K$\alpha$ line
observed in eclipse is produced in the cool regions of the extended stellar
wind.  On the other hand, we find that this same model cannot explain the
observed intensity of the K$\alpha$ line in the other two phases.  The
equivalent widths from the simulations are only 4~eV and 11~eV for phases 0.25
and 0.50, whereas the observed values are 51~eV and 116~eV, respectively.

As an obvious source of additional line production regions, we first consider
the surface of the companion star.  Using Monte Carlo simulations and adopting
the geometrical parameters listed in Table~\ref{tbl:simparams}, we obtain
reflected iron K$\alpha$ equivalent widths of 29~eV and 65~eV during phase
0.25 and phase 0.50, respectively, assuming that the metal abundance of the
companion star is also 0.75 times the cosmic value.  Therefore, the stellar
wind and the companion surface combined yield equivalent widths of 33~eV
(phase 0.25) and 76~eV (phase 0.50).  The observed values are still in excess
these values by 18~eV and 40~eV at phase 0.25 and 0.5, respectively.

We note the presence of excess absorption of
$N_{\mathrm{H}}$~=~1.7~$\times$~10$^{23}$~cm$^{-2}$ observed during phase 0.5
relative to that at phase 0.25.  This suggests that a distribution of cold
localized material lies along the line of centers somewhere behind the neutron
star as viewed from the companion.  We therefore expect that at least some
amount of Fe-K line photons are produced in this cloud.  This has already been
pointed out by \cite{inoue85} using the {\it Tenma} data.

\begin{figure}
\epsscale{0.5}
\plotone{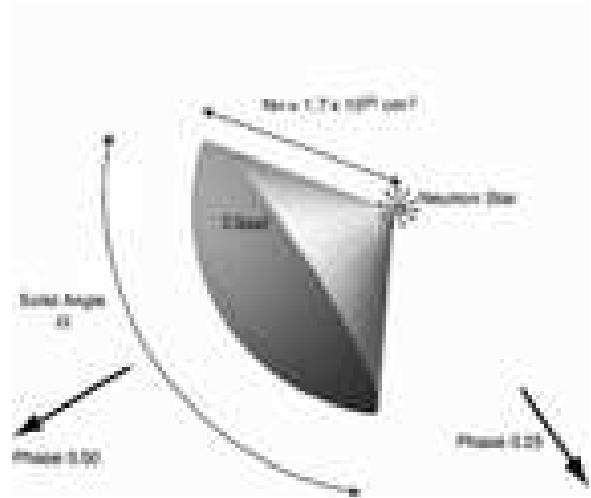}
\caption{Geometry of the cold cloud used in the Monte Carlo simulation to
calculate the iron K$\alpha$ equivalent width.}
\label{fig:partial_cloud}
\end{figure}

In order to estimate the contribution of the cloud to the iron line flux, we
have performed a Monte Carlo simulation of a partially covered cloud
irradiated by an X-ray source.  The geometrical configuration is shown
schematically in Fig.~\ref{fig:partial_cloud}.  The thickness of the cloud
is assumed to be $N_{\mathrm{H}} = 1.7 \times 10^{23}$~cm$^{-2}$ and we
assume a uniform density.  The cloud is assumed to be neutral and its metal
abundance is set to 0.75 cosmic.  The resulting emission spectra as functions
of the orbital phase with the solid angle as the free parameter are computed
through the same procedures as in the previous simulations, and the simulated
Fe K$\alpha$ line equivalent widths are measured at the three orbital phases.

\begin{figure}
\epsscale{0.6}
\plotone{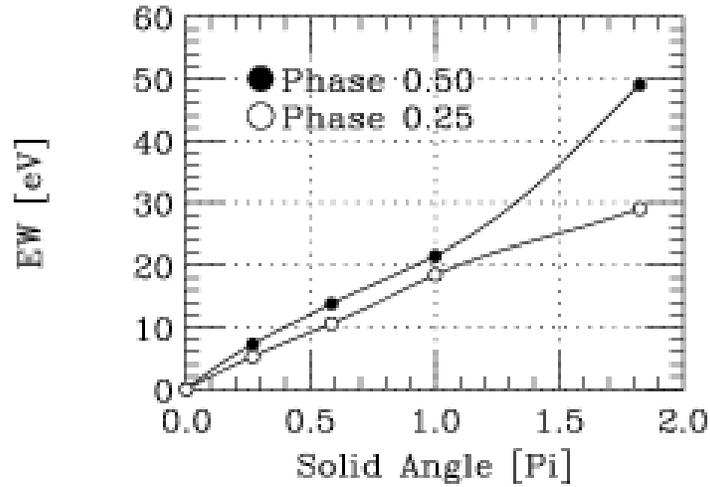}
\caption{Relation between the cloud's solid angle and the equivalent width
of the iron K$\alpha$ line calculated with a Monte Carlo simulation.  The
thickness of the cloud is $N_{\mathrm{H}} = 1.7 \times 10^{23}$~cm$^{-2}$ and
assumed to be at a uniform density.  Open and filled circles show the
predicted equivalent width as observed at phases 0.25 and 0.50, respectively.}
\label{fig:eqw_cover}
\end{figure}

Figure~\ref{fig:eqw_cover} shows the relation between the solid angle
subtended by the cloud and the iron line equivalent width.  As can be seen
from this figure, the missing equivalent widths outside of eclipse (18~eV at
phase 0.25 and 40~eV at phase 0.50) can be explained if the cloud covers
25--40~\% of the sky as viewed from the neutron star.  Although this
geometrical configuration may be oversimplified, its general characteristics
are consistent with the fact that the spectrum observed at phase 0.25 does not
exhibit heavy absorption and the fact that soft X-ray lines are still observed
during phase 0.5.  \cite{sato86a} reported a gradual increase of average
absorption column density between orbital phase $\sim 0.2$ and $\sim 0.9$ from
$N_{\mathrm{H}}$~$\simeq$~1~$\times$~10$^{22}$~cm$^{-2}$ to
$N_{\mathrm{H}}$~$\simeq$~3~$\times$~10$^{23}$~cm$^{-2}$ observed in the {\it
  Tenma} data of Vela X-1.

The enhanced absorption at orbital phase 0.50, and the localized cloud
responsible for the iron line production can be attributed to an ``accretion
wake'', which is produced by a combination of several physical effects,
including gravitational, rotational, and radiation pressure forces, and X-ray
heating.  Numerical simulations by \cite{sawada86}, \cite{blondin90},
\cite{blondin91} and \cite{blondin94} exhibit such structures trailing the
accreting neutron star.

\subsection{The Velocity Field of the Stellar Wind}

\subsubsection{Mismatch between observation and simulation}

When the observed line profiles of a few of the brightest lines are compared
with the output of the Monte Carlo simulations, we find that there is a
mismatch in the average velocity shift at both phase 0.5 and eclipse.
Figure~\ref{fig:comparison_doppler} shows comparison of the observed
\ion{Si}{14} and \ion{Mg}{12} Ly$\alpha$ lines with the simulations convolved
with the instrument response.  Although the sign of the shifts are correct,
the simulations seem to predict a larger velocity shift than seen in in the
data.  Figure~\ref{fig:dopplershift} shows the velocity shifts for all of the
bright highly ionized emission lines detected in the spectrum.  The average
velocity shifts ($\Delta v$) observed between phase 0.5 and eclipse lie range
in $\sim$~300--600~km~s$^{-1}$ whereas the simulations predict $\Delta v
\sim$~1000~km~s$^{-1}$, which is larger approximately by a factor of two.

\subsubsection{Interaction between X-rays and the stellar wind}

One of the most obvious candidate mechanisms that is capable of reducing the
wind velocity is X-ray photoionization.  In the \cak -model of isolated OB
stars, a stellar wind is driven by the radiation force on the large number of
resonance lines that lie in the UV.  In particular, the dominant ions that are
responsible for this are the L-shell ions of C, N and O, which have strong
valence shell transitions in the UV range and have large abundances compared
to the other metals.

In \hmxb s, there are regions in the stellar wind where X-ray photoionization
destroys the L-shell ions of C, N, and O into K-shell ions and, therefore,
reduce the amount of radiative force.  This effect is maximized near the
neutron star where the X-ray flux is high and which also happens to coincide
with the region where the wind would otherwise be accelerated in the absence
of the X-ray source.  As a result, the stellar wind velocity within this
region should be lower than that predicted by the \cak -model.  The effect of
X-ray irradiation on line-driven stellar winds in \hmxb\ systems has been
studied theoretically by \cite{macgregor82} and \cite{masai84}, and they have
also been seen in UV spectral observations \citep{kaper93}.

\subsubsection{One dimensional calculation of the velocity structure \label{sec:1dcalcv}}

In order to investigate how the stellar wind velocity is affected due to X-ray
ionization, we have performed a simple one-dimensional calculation to estimate
the velocity profile along the line of centers.  The extension of this
calculation to three-dimensions is required to self-consistently incorporate
into the Monte Carlo simulator and predict the emission line spectra, which is
beyond the scope of this paper.  Nevertheless, as we show below, this estimate
in 1-D is still useful for understanding semi-quantitatively the mismatch
between the observed and predicted velocity shifts.

\begin{figure}
\epsscale{0.8}
\plotone{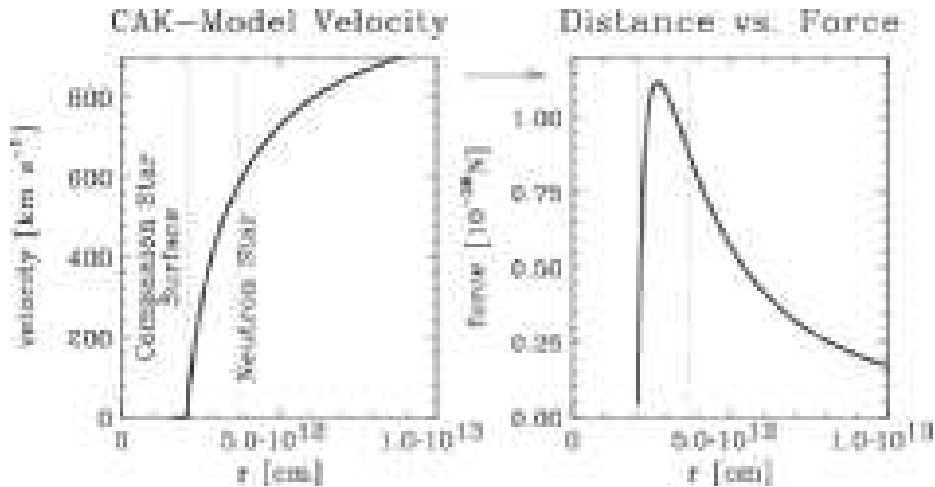}
\caption{The velocity profile of the CAK-model stellar wind with
$v_{\infty} = 1100$~km~s$^{-1}$ and $\beta = 0.8$ (left) and the force profile
(right).  The vertical dotted lines designate the location of the stellar
surface and the neutron star.  The force is numerically derived from the
CAK-model velocity structure.  Note that the force is most dominant in the
region between the neutron star and the companion stellar surface.}
\label{fig:cakmodel_v_f_r}
\end{figure}

In this calculation, we make two simplifying assumptions.  We first assume
that the wind is driven radiatively by UV resonance absorption and that there
are no other relevant forces that provide additional acceleration.  The other
assumption is that the force is proportional to the local number densities of
C, N and O ions with more than one electron in the L-shell (i.e., Li-like and
lower charge states).  Figure~\ref{fig:cakmodel_v_f_r} shows the velocity as a
function of distance along the line of centers in the \cak -model and the
radiative force, which is numerically derived from the \cak -model velocity
structure.  As can been seen from these figures, the radiation force peaks in
the region between the companion star surface and the location of the neutron
star.  This is, however, the region where most of the X-ray lines are produced
in this system.  Therefore, the acceleration of the stellar wind in this
region must be reduced since the C, N, and O ions inside this region are
primarily H-like, He-like, or fully-stripped, and they alone cannot provide
the force required to maintain the acceleration.

\begin{figure}
\epsscale{0.6}
\plotone{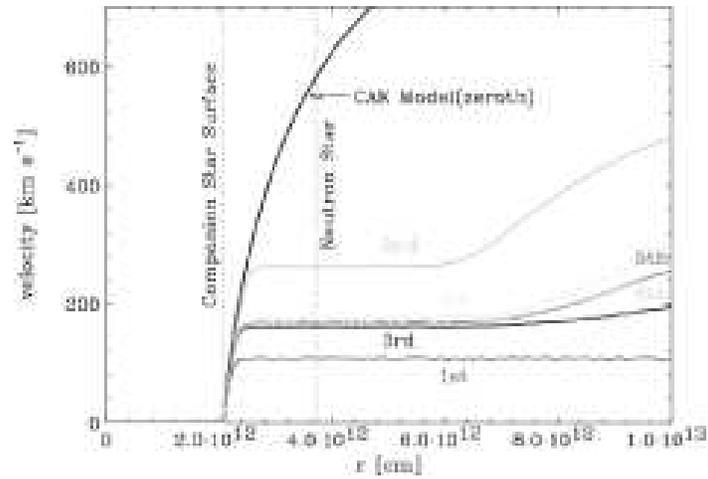}
\caption{The results of a 1-D calculation of the stellar
wind velocity along the axis between the companion star and the neutron star
accounting for X-ray photoionization as described in \S~\ref{sec:1dcalcv}.  
The mass loss
rate is assumed to be $1.5 \times 10^{-6} M_{\sun} \mathrm{yr}^{-1}$.  The
flattening of the velocity profile is due to X-ray photoionization destroying
the ions that are capable of absorbing in the UV.}
\label{fig:velocity_calc}
\end{figure}

We estimate the magnitude of this effect as follows.  We start with the \cak
velocity profile, which has an implicit assumption about the force as a
function of distance, and can calculate the density everywhere in the wind.
This can then be used to derive the ionization structure and, in particular,
the distribution of C, N, and O ions, which provides an estimate of the new
force and hence the velocity profile.  This process can be repeated and
iterated to find an ionization structure that is self-consistent with the
velocity profile through the following steps,
\begin{enumerate}
  \item start from the velocity structure ($v_{1}'(r) = v_{0}(r)$ (\cak -model)),
  	and, calculate a force structure ($f_{1}(r)$), then, obtain $v_{1}(r)$.
  \item start from $v_{2}'(r) = v_{1}(r)$, and, calculate $f_{2}(r)$, then 
  obtain $v_{2}(r)$.
  \item start from $v_{3}'(r) = (v_{1}(r) + v_{2}(r))/2$, and, calculate 
  $f_{3}(r)$, then obtain $v_{3}(r)$.
  \item start from $v_{4}'(r) = v_{3}(r)$, and, calculate $f_{4}(r)$, then 
  obtain $v_{4}(r)$.
  \item start from $v_{5}'(r) = (v_{3}(r) + v_{4}(r))/2$, and, calculate 
  $f_{5}(r)$, then obtain $v_{5}(r)$.
  \item start from $v_{6}'(r) = v_{5}(r)$, and, calculate $f_{6}(r)$, then 
  obtain $v_{6}(r)$.  
\end{enumerate}
The computed velocity structures ($v_{0}(r)$, $v_{1}(r)$, $v_{2}(r)$,
$v_{3}(r)$, $v_{4}(r)$, $v_{5}(r)$ and $v_{6}(r)$) are plotted in
Fig.~\ref{fig:velocity_calc}.  As shown, the solution has converged after the
5th iteration.  In this simple model, the stellar wind velocity at the point
of the neutron star is only $\sim$~180~km~s$^{-1}$ compared to
$\sim$~570~km~s$^{-1}$ predicted by the \cak -model.  This reduction in
velocity of $\sim$~390~km~s$^{-1}$ is remarkably similar to the
observed velocity offset between the model and data observed during phase 0.5,
where most of the emission is believed to originate from near the neutron
star.

Our one dimensional calculations confirm that the photoionization by the X-ray
radiation can make the stellar wind near the neutron star slower by a factor
of 2--3 compared to an undisturbed \cak\ wind.  This results are in agreement
with the observed Doppler shifts in the emission lines. The X-ray radiation of
the neutron star is disturbing the flow of the \cak -model stellar wind in the
Vela~X-1 system, and then, the velocity of the stellar wind in the highly
photoionized region becomes slower than that estimated from \cak -model. 

In order to understand the flow of stellar wind precisely and obtain 
the X-ray spectrum from the photoionized stellar wind,
further considerations should be needed.
A three dimensional calculation of the electromagnetic hydrodynamics
would be required for a fully quantitative analysis.
The lower velocity of the stellar wind due to the X-ray photoionization
leads to a density increase and changes of emission line intensities.
In that case, an adjustment of the mass loss rate could be needed to 
reproduce the observed spectra. Though the mass loss rate is proportional to
the density and the velocity, how large the adjustment is required is not
simple. The neutron star goes around the companion star, and, the photoionized
region moves in association with the orbital motion of the neutron star. 
It takes about 0.5 days for the stellar wind to reach the position of the 
neutron star from the surface of the companion star assuming the \cak -model.
It is not negligible in comparison to the orbital period of the neutron star 
(8.96 days), and, calculations including the time evolution due to the 
orbital motion are required. Additionally, gas pressure forces not only
in the radial directions but along the circumferential directions are
important since the stellar wind flow is not spherically symmetry any more.

\section{Summary and Conclusions}

\begin{figure}
\epsscale{0.9}
\plotone{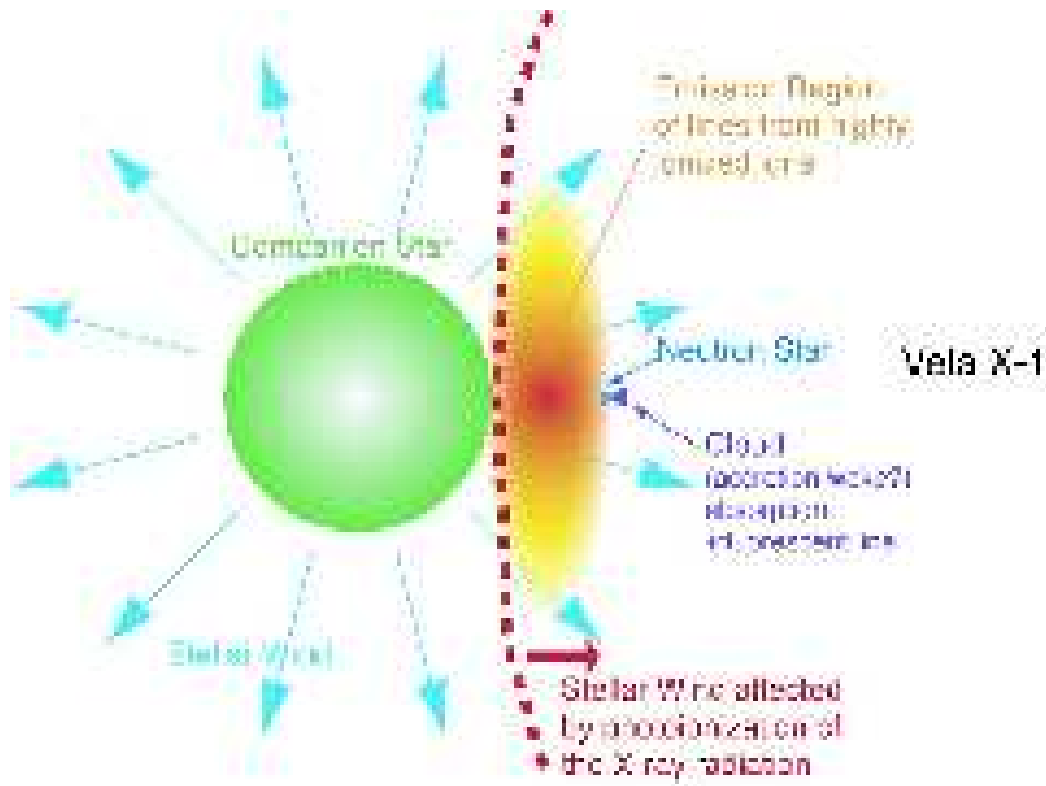}
\caption{A conceptual picture of the Vela X-1 system obtained from the present
analysis.}
\label{fig:hmxb_illust}
\end{figure}

Vela X-1 was observed by the \Chandra\ \hetgs\ at three distinct orbital phase
ranges.  In order to understand the geometry and the dynamics of the X-ray
irradiated stellar wind in this system, we have developed a Monte Carlo
spectral simulator to predict the emission line spectrum for a given set of
stellar wind parameters.  The model spectra were then compared to the observed
spectra to construct a self-consistent picture that explains the global
properties of the X-ray spectra.  The qualitative aspects of this model and
its various components are illustrated in Fig~\ref{fig:hmxb_illust}.

The X-ray emission line intensities of highly ionized H- and He-like ions are
well-reproduced by a \cak -model of a stellar wind with a mass loss rate of
(1.5--2.0)~$\times 10^{-6} M_{\sun} \mathrm{yr}^{-1}$ and a terminal velocity
of 1100~km~s$^{-1}$ irradiated by the X-ray continuum radiation from the
neutron star.  Through comparisons of the spectra observed at eclipse and at
phase 0.50, we find that most of the X-ray recombination emission lines are
produced in the region between the neutron star and the companion star (see
Fig.~\ref{fig:hmxb_illust}) where both the gas density and the local X-ray
flux are high.  This region is occulted by the companion star during the
eclipse phase, which explains the relative strengths of the emission line
intensities during phase 0.50 and eclipse.  The same set of model parameters
also nicely explains the lack of obvious discrete spectral features observed
at phase 0.25.

The properties of the intense iron fluorescent lines, which are observed
throughout the binary orbit, are sensitive to the distribution of cold
material relative to the location of the neutron star.  Although the iron
fluorescent line intensity observed during eclipse can be explained entirely
by fluorescence in the extended stellar wind, the line intensities during the
other two phases are far brighter than predicted by the same model.  We find
that, in addition to the stellar wind and the X-ray irradiated surface of the
companion star, a cold cloud partially covering the neutron star is required
to explain the data.  From the quantitative analysis, we find that a cloud
with a thickness of $N_{\mathrm{H}} = 1.7 \times 10^{23}$~cm$^{-2}$ and covers
25--40~\% of the solid angle viewed from the neutron star, explains both the
iron K$\alpha$ line intensities as well as the excess column density observed
during phase 0.5.

The one observed feature that this model fails to reproduce is the velocity
shift of the highly-ionized recombination emission lines, which are observed
to be smaller in magnitude by several hundred $\rm{km~s}^{-1}$ compared to the
predictions.  Using a simple 1-D model, we find that destruction of
UV-absorbing ions by X-ray photoionization can explain this reduction in
velocity. However, the fully self-consistent 3-D model including the X-ray 
photoionization effect still remains an issue to be resolved.
A full 3-D electromagnetic hydrodynamics calculation including time dependent effects
would be required for the further study.

\begin{figure}
\epsscale{0.7}
\plotone{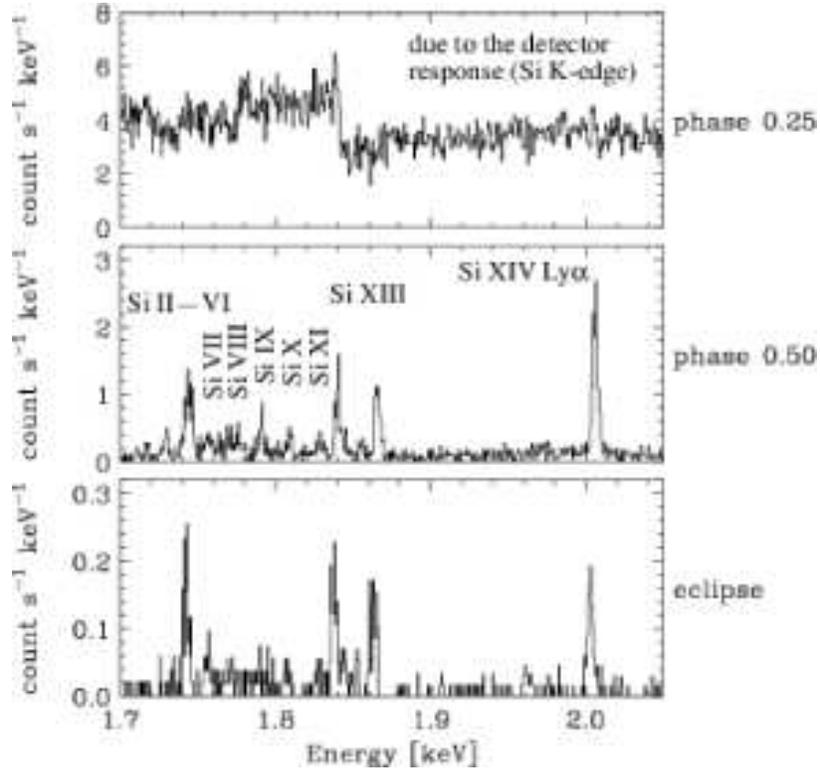}
\caption{The spectra of the Si K line complex observed at each orbital phase
-- phase 0.25 (top), phase 0.50 (middle), eclipse (bottom).  The phase 0.25
   spectrum is dominated by the intense continuum with possible evidence for
   weak absorption lines.  The sharp drop at $E = 1.84~\rm{keV}$ is due Si K
   photoelectric absorption in the detector.  At phases 0.50 and eclipse,
   emission lines from a wide range of charge states are seen.}
\label{fig:si_line_spec}
\end{figure}

In addition, there are much more things to be done with this data set. These include
study of the Compton shoulder in the iron K line and the ratio of He-like triplet lines,
which constrain the physical conditions in the emission region.
There is also much more to be improved in the Monte Carlo code. For example, the code for
ions from Li-like to near neutral is needed for a detailed study of 
the Si K line complex shown in Fig.~\ref{fig:si_line_spec}.
Such topics should be subjects of the next paper.


\begin{deluxetable}{crccc}
\tablecolumns{5}
\tablewidth{0pt}
\tablecaption{Summary of Vela X-1 Observations \label{tbl:velax1obslist}}
\tablehead{
  Label & OBSID & Start Date & Orbital Phase & Exposure (sec)
}
\startdata
0.25    & 1928 & 2001-02-05 05:29:55 & 0.237 -- 0.278 & 29570 \\
0.50    & 1927 & 2001-02-07 09:57:17 & 0.481 -- 0.522 & 29430 \\
eclipse & 1926 & 2001-02-11 21:20:17 & 0.980 -- 0.093 & 83150 \\
\enddata
\end{deluxetable}

\begin{deluxetable}{cccccc}
\tablecolumns{6}
\tablewidth{0pt}
\tablecaption{Properties of continuum spectra derived from spectral fits.
\label{tbl:cont_fit}}
\tablehead{
Orbital & $N_\mathrm{H}$\tablenotemark{a} & Photon Index & Observed Flux  & Luminosity\tablenotemark{b} & $\chi^2$/ d.o.f. 
 \\
Phase & ($10^{22}$~cm$^{-2}$) & & (erg~cm$^{-2}$~s$^{-1}$) & (erg~s$^{-1}$) & }
\startdata
0.25 & 1.45~$\pm$~0.03 & 1.01~$\pm$~0.01 & 3.0~$\times$~10$^{-9}$ & 1.6~$\times$~10$^{36}$ & 2872./ 3318 \\
0.50 & 18.5~$\pm$~0.3 & 1.01 (fixed) & 1.5~$\times$~10$^{-9}$ &
1.6~$\times$~10$^{36}$ & 1000./ 1051 \\
\enddata

\tablecomments{Fitting regions are 1.0--10.0~keV and 3.0--10.0~keV from HEG
for 0.25 and 0.50 orbital phases, respectively.
The iron K-line
region (6.3--6.5~keV) is excluded.
Errors correspond to 90~\% confidence levels.}
\tablenotetext{a}{The metal abundance is assumed to be 0.75 cosmic.}
\tablenotetext{b}{0.5--10.0~keV luminosity corrected for absorption.}
\end{deluxetable}

\begin{deluxetable}{ccccc}
\tablecolumns{5}
\tablewidth{0pt}
\tablecaption{Derived Parameters of emission lines in the 0.5 orbital phase 
spectrum \label{tbl:050_line_fit}}
\tablehead{
Energy & Sigma & Flux\tablenotemark{a} & Candidate & Line Shift \\
(keV) & (eV) & ($10^{-5}$ photon cm$^{-2}$ s$^{-1}$) &      & (km s$^{-1}$)
}
\startdata
$3.6905^{+0.0022}_{-0.0009}$      & $0.4_{-0.4}^{+3.4}$       & 8.8{\scriptsize $\pm$3.1}  & Ca II--XII K$\alpha$ & \\ 
2.622{\scriptsize $\pm$0.001}     & $4.0_{-1.2}^{+1.5}$       & $17.7_{-3.3}^{+3.2}$       & S XVI Ly$\alpha$ & \\
$2.46197^{+0.00076}_{-0.00087}$   & $0.6_{-0.6}^{+1.1}$       & $6.8_{-1.7}^{+1.8}$        & S XV r & \\
$2.31112^{+0.00072}_{-0.00042}$   & $0.2_{-0.2}^{+1.9}$       & $10.3_{-2.0}^{+2.2}$       & S IV--VIII K$\alpha$ & \\
2.00634{\scriptsize $\pm$0.00016} & $1.2_{-0.3}^{+0.2}$       & $23.8_{-1.5}^{+1.4}$       & Si XIV Ly$\alpha$ & +127{\scriptsize $\pm$24} \\
1.86614{\scriptsize $\pm$0.00022} & 1.4{\scriptsize $\pm$0.3} & $13.6_{-1.1}^{+1.2}$       & Si XIII r & $+183_{ -35}^{ +36}$ \\
$1.85537^{+0.00055}_{-0.00054}$   & $0.8_{-0.8}^{+1.0}$       & $3.1_{-0.8}^{+0.7}$        & Si XIII i & $+257_{ -87} ^{+89}$\\
$1.84136^{+0.00026}_{-0.00027}$   & 2.2{\scriptsize $\pm$0.3} & 14.1{\scriptsize $\pm$1.2} & Si XIII f & $+311_{ -44}^{ +42}$\\
1.74447{\scriptsize $\pm$0.00026} & 2.4{\scriptsize $\pm$0.3} & $14.6_{-1.0}^{+1.2}$       & Si II--VI K$\alpha$        & \\
$1.72998^{+0.00035}_{-0.00036}$   & $0.4_{-0.4}^{+0.7}$       & 3.4{\scriptsize $\pm$0.6}  & Al XIII Ly$\alpha$? & \\
$1.59900^{+0.00067}_{-0.00068}$   & $0.7_{-0.7}^{+1.0}$       & $1.3_{-0.4}^{+0.6}$        & Al XII r ?& \\
1.57976{\scriptsize $\pm$0.00056} & 2.0{\scriptsize $\pm$0.6} & 3.9{\scriptsize $\pm$0.7}  & Mg XI & \\
$1.55231^{+0.00053}_{-0.00056}$   & $0.6_{-0.6}^{+0.8}$       & $1.8_{-0.5}^{+0.6}$        & Fe XXIV ? & \\
1.47282{\scriptsize $\pm$0.00014} & 1.2{\scriptsize $\pm$0.2} & 23.5{\scriptsize $\pm$1.8} & Mg XII Ly$\alpha$ & $+102_{ -29}^{ +28}$\\
$1.35279^{+0.00017}_{-0.00016}$   & 1.0{\scriptsize $\pm$0.2} & 14.7{\scriptsize $\pm$1.5} & Mg XI r & $+120_{ -36}^{ +37}$\\
$1.34346^{+0.00021}_{-0.00022}$   & 0.7{\scriptsize $\pm$0.3} & $7.3_{-1.0}^{+1.2}$        & Mg XI i & $+80_{ -49}^{+57}$\\
1.33213{\scriptsize $\pm$0.00023} & 1.0{\scriptsize $\pm$0.3} & $9.1_{-1.2}^{+1.4}$        & Mg XI f & $+230_{ -52}^{ +51}$\\
$1.30808^{+0.00035}_{-0.00042}$   & $1.1_{-0.5}^{+0.4}$       & $5.0_{-1.0}^{+1.1}$        & Fe XXI ? & \\
$1.27783^{+0.00028}_{-0.00027}$   & 1.1{\scriptsize $\pm$0.3} & $8.4_{-1.2}^{+1.4}$        & Ne X Ly$\gamma$ & \\
1.21160{\scriptsize $\pm$0.00021} & $1.0_{-0.3}^{+0.2}$       & $13.3_{-1.8}^{+1.9}$       & Ne X Ly$\beta$ & \\
$1.12732^{+0.00048}_{-0.00049}$   & $1.5_{-0.4}^{+0.5}$       & $7.4_{-1.7}^{+1.9}$        & Ne IX & \\
$1.07432^{+0.00018}_{-0.00035}$   & $0.1_{-0.1}^{+0.6}$       & $5.8_{-1.7}^{+1.9}$        & Ne IX & \\
$1.02242^{+0.00011}_{-0.00012}$   & $0.6_{-0.1}^{+0.2}$       & $44.2_{-5.1}^{+5.5}$       & Ne X Ly$\alpha$ & $+182_{ -35}^{+32}$\\
0.92246{\scriptsize $\pm$0.00027} & 0.8{\scriptsize $\pm$0.3} & $28.9_{-6.5}^{+7.5}$       & Ne IX r & $+149_{ -89}^{ +86}$\\
$0.91625^{+0.00033}_{-0.00032}$   & $1.4_{-0.2}^{+0.3}$       & $49.1_{-8.8}^{+9.7}$       & Ne IX i & $+475_{-104}^{ +108}$\\
$0.90556^{+0.00051}_{-0.00055}$   & $1.8_{-0.4}^{+0.5}$       & $37.1_{-8.3}^{+9.4}$       & Ne IX f & $+165_{-181}^{ +169}$\\
\enddata

\tablenotetext{a}{Inter stellar gas absorption is corrected. The hydrogen
column density of 6~$\times$~ 10$^{21}$~cm$^{-2}$ is assumed, corresponding to
the density of 1~H~cm$^{-3}$ and the distance of 1.9~kpc.}
\tablecomments{Errors correspond to 90~\% confidence level.}

\end{deluxetable}

\begin{deluxetable}{ccccc}
\tablecolumns{5}
\tablewidth{0pt}
\tablecaption{Derived Parameters of emission lines in the eclipse phase 
spectrum \label{tbl:000_line_fit}}
\tablehead{
Center Energy & Sigma & Flux\tablenotemark{a} & Candidate & Line Shift \\
(keV) & (eV) & ($10^{-6}$photon cm$^{-2}$ s$^{-1}$) &  & (km s$^{-1}$)
}
\startdata
$3.69431^{+0.0040}_{-0.0043}$    & $8.6_{-1.2}^{+4.3}$  & $9.3_{-2.7}^{+3.0}$  & Ca II--XII K$\alpha$ &  \\
$2.95657^{+0.00119}_{-0.0015}$   & $0.0_{-0.0}^{+3.4}$  & $4.9_{-2.0}^{+2.4}$  & Ar VI--IX K$\alpha$ & \\
$2.61857^{+0.0009}_{-0.0076}$    & $0.6_{-0.6}^{+2.0}$  & $13.8_{-3.0}^{+3.5}$ & S XVI Ly$\alpha$ & \\
2.31035{\scriptsize $\pm$0.0010} & 2.2{\scriptsize $\pm$1.3} & $17.6_{-3.9}^{+4.7}$ & S IV--VIII K$\alpha$ & \\
2.00339{\scriptsize $\pm$0.00034}& 1.7{\scriptsize $\pm$0.4} & $23.2_{-2.5}^{+2.7}$ & Si XIV Ly$\alpha$ & $-$314{\scriptsize $\pm$51} \\
$1.86299^{+0.00031}_{-0.00032}$  & 1.6{\scriptsize $\pm$0.3} & $23.4_{-2.6}^{+2.8}$ & Si XIII r & $-323_{-52}^{+50}$\\
$1.85271^{+0.00086}_{-0.00017}$  & $0.0_{-0.0}^{+1.1}$ & $2.8_{-1.1}^{+1.3}$  & Si XIII i & $-173_{-28}^{+139}$ \\
$1.83924^{+0.00045}_{-0.00046}$  & $2.9_{-0.4}^{+0.5}$ & $21.1_{-2.3}^{+2.5}$ & Si XIII f & $-34_{-75}^{+73}$\\
$1.74247^{+0.00029}_{-0.00030}$  & 1.7{\scriptsize $\pm$0.3} & $19.6_{-2.2}^{+2.4}$ & Si II--VI K$\alpha$ & \\
$1.65752^{+0.0010}_{-0.00055}$   & $0.1_{-0.1}^{+1.4}$  & $2.8_{-0.9}^{+1.2}$  & Mg XI or  Fe XXIII & \\
$1.57719^{+0.00057}_{-0.00060}$  & $1.3_{-0.6}^{+0.8}$  & $6.0_{-1.4}^{+1.6}$  & Mg XI              & \\
$1.47068^{+0.00025}_{-0.00024}$  & $1.2_{-0.2}^{+0.3}$  & $26.3_{-3.4}^{+3.6}$ & Mg XII Ly$\alpha$ & $-334_{-49}^{+51}$\\
1.35057{\scriptsize $\pm$0.00031}& $1.5_{-0.3}^{+0.4}$  & $25.8_{-3.5}^{+3.9}$ & Mg XI r & $-$373{\scriptsize $\pm$69}\\
$1.34238^{+0.00069}_{-0.00068}$  & 1.2{\scriptsize $\pm$0.8} & $6.3_{-2.0}^{+2.4}$  & Mg XI i & $-161_{-152}^{+154}$\\
$1.33122^{+0.00049}_{-0.00050}$  & $2.3_{-0.4}^{+0.5}$  & $20.3_{-3.3}^{+3.7}$ & Mg XI f & $+25_{-114}^{+110}$ \\
$1.30640^{+0.00061}_{-0.00066}$  & $1.3_{-0.6}^{+0.7}$  & $6.6_{-2.0}^{+2.5}$  & Fe XXI ?              & \\
$1.27592^{+0.00044}_{-0.00045}$  & $0.9_{-0.9}^{+0.5}$  & $8.7_{-3.1}^{+2.7}$  & Ne X Ly$\gamma$       & \\
1.23618{\scriptsize $\pm$0.0011} & $1.6_{-0.9}^{+1.2}$  & $5.9_{-2.8}^{+3.1}$  & Fe XX ?               & \\
$1.20980^{+0.00053}_{-0.00050}$  & $1.3_{-0.3}^{+0.5}$  & $14.3_{-3.6}^{+4.1}$ & Ne X Ly$\beta$        & \\
$1.07241^{+0.00035}_{-0.00036}$  & 0.8{\scriptsize $\pm$0.4} & $19.1_{-5.3}^{+6.2}$ & Ne IX                 & \\
$1.02075^{+0.00020}_{-0.00019}$  & 1.0{\scriptsize $\pm$0.2} & $85_{-13}^{+15}$ & Ne X Ly$\alpha$ & $-308_{-56}^{+59}$\\
$0.92077^{+0.00039}_{-0.00038}$  & $0.9_{-0.3}^{+0.4}$  & $53_{-16}^{+21}$    & Ne IX r  & $-400_{-124}^{+126}$ \\
0.91493{\scriptsize $\pm$0.00052}& $1.1_{-0.3}^{+0.4}$  & $47_{-17}^{+20}$    & Ne IX i  & $+40_{-170}^{+171}$\\
$0.90406^{+0.00028}_{-0.00027}$  & 0.8{\scriptsize $\pm$0.3} & $90_{-22}^{+26}$ & Ne IX f  & $-333_{-91}^{+94}$ \\
\enddata

\tablenotetext{a}{Inter stellar gas absorption is corrected. The hydrogen
column density of 6~$\times$~ 10$^{21}$~cm$^{-2}$ is assumed, corresponding to
the density of 1~H~cm$^{-3}$ and the distance of 1.9~kpc.}
\tablecomments{Errors correspond to 90~\% confidence level.}

\end{deluxetable}

\begin{deluxetable}{rcc}
\tablecolumns{3}
\tablewidth{0pt}
\tablecaption{Comparison between lines of the phase 0.50 and of the eclipse.\label{tab:comparison_lines}}
\tablehead{
line & Flux(0.50)/Flux(eclipse) & $\Delta v$ (km~s$^{-1})$
}
\startdata
Si XIV Ly$\alpha$ & 10.3~$\pm$~1.3  & 441~$\pm$~56 \\
Si XIII resonance & 5.81~$\pm$~0.83 & 506~$\pm$~62 \\
Si XIII forbidden & 6.68~$\pm$~0.95 & 345~$\pm$~86 \\
Mg XII Ly$\alpha$ & 8.94~$\pm$~1.37 & 436~$\pm$~58 \\
Mg XI resonance &   5.70~$\pm$~1.00 & 493~$\pm$~78 \\
Mg XI forbidden &   4.50~$\pm$~1.01 & 205~$\pm$~123 \\
Ne X Ly$\alpha$ &   5.18~$\pm$~1.04 & 490~$\pm$~60 \\
Ne IX resonance &   5.27~$\pm$~2.14 & 549~$\pm$~153 \\
Ne IX forbidden &   4.13~$\pm$~1.49 & 498~$\pm$~198 \\
\enddata
\end{deluxetable}

\begin{deluxetable}{ccccc}
\tablecolumns{5}
\tablewidth{0pt}
\tablecaption{Derived Parameters of the Fe K$\alpha$ line.\label{tab:fekalpha}}
\tablehead{
Orbital Phase & Energy & Sigma & Flux & EW \\
              & (keV)  & (eV)  & ($10^{-4}$photon cm$^{-2}$ s$^{-1}$) & (eV) 
}
\startdata
0.00 & 6.3958$\pm$0.0022          & $7.2_{-5.3}^{+3.9}$ & 1.7$\pm$0.2 & 844 \\
0.25 & $6.3992_{-0.0005}^{+0.0018}$ & $0.0_{-0.0}^{+7.4}$ & $19.2_{-1.4}^{+1.5}$ & 51 \\
0.50 & $6.3965_{-0.0012}^{+0.0011}$ & $11.0_{-1.9}^{+1.8}$& $34.0_{-1.9}^{+1.8}$ & 116 \\
\enddata

\tablecomments{Only HEG data is used. The fitting model is single gaussian.}
\tablecomments{Errors correspond to 90~\% confidence level.}

\end{deluxetable}

\begin{deluxetable}{lrl}
\tablecolumns{3}
\tablewidth{0pt}
\tablecaption{Adopted parameters for the Vela X-1 simulation \label{tbl:simparams}} 
\tablehead{
Parameter & Value & Reference
}
\startdata
Geometry \nodata & &  \\
\  Binary Separation $D$ & 53.4~$R_{\sun}$& \cite{kerkwijk95} \\
\  Companion Star Radius $R_{*}$ & 30.0~$R_{\sun}$ & \cite{kerkwijk95} \\ 
Stellar Wind \nodata & & \\
\  Velocity Structure $v(r)$ & $v_{\infty} (1 - R_{*}/r)^{\beta}$ & 
\cite{castor75} \\
\  Terminal Velocity $v_{\infty}$ & 1100~km~s$^{-1}$ & \cite{prinja90} \\
\  $\beta$ & 0.80 & \cite{pauldrach86} \\
\  Metal Abundance ($Z > 2$) & 0.75 cosmic & for typical OB-stars \\
                             &            &      \citep{bord76} \\
X-ray Radiation \nodata & & \\
\  Luminosity(0.5--20~keV) & 3.5~$\times$~10$^{36}$~erg~s$^{-1}$ & 
This observation \\
& & (1.6~$\times$~10$^{36}$~erg~s$^{-1}$ (0.5--10~keV)) \\
\  Spectrum Shape & Power-Law ($\Gamma$~=~1.0) & This observation \\
\  Energy Range & 13.6~eV -- 20.0~keV & \\
\enddata
\end{deluxetable}

\clearpage

\clearpage

\end{document}